\documentclass[twocolumn, times, tighten,twocolappendix]{aastex631}

\usepackage{graphicx,multirow,hyperref,url,color,xspace}
\usepackage{amsmath,amssymb}

\newcommand{\msun}{{\rm M}_{\sun}}

\newcommand{\integral}{{\textit{INTEGRAL}}\xspace}
\newcommand{\source}{{MAXI J1820+070}\xspace}

\newcommand{\appropto}{\mathrel{\vcenter{
  \offinterlineskip\halign{\hfil$##$\cr
    \propto\cr\noalign{\kern2pt}\sim\cr\noalign{\kern-2pt}}}}}

\defcitealias{Tetarenko21}{T21}
\newcommand{\tet}{{\citetalias{Tetarenko21}}\xspace}
\defcitealias{Zdziarski19b}{Z19}
\newcommand{\zdz}{{\citetalias{Zdziarski19b}}\xspace}
\defcitealias{ZSPT15}{Z15}
\newcommand{\core}{{\citetalias{ZSPT15}}\xspace}

\begin{document}

\title{Jet Parameters in the Black-Hole X-Ray Binary MAXI J1820+070}
\shorttitle{Jet parameters}

\author{Andrzej A. Zdziarski}
\affiliation{Nicolaus Copernicus Astronomical Center, Polish Academy of Sciences, Bartycka 18, PL-00-716 Warszawa, Poland; \href{mailto:aaz@camk.edu.pl}{aaz@camk.edu.pl}}
\author{Alexandra J. Tetarenko}
\altaffiliation{NASA Einstein Fellow}
\affiliation{East Asian Observatory, 660 N. A'oh\={o}k\={u} Place, University
Park, Hilo, Hawaii 96720, USA}
\affiliation{Department of Physics and Astronomy, Texas Tech University, Lubbock, Texas 79409-1051, USA}
\author{Marek Sikora}
\affiliation{Nicolaus Copernicus Astronomical Center, Polish Academy of Sciences, Bartycka 18, PL-00-716 Warszawa, Poland; \href{mailto:aaz@camk.edu.pl}{aaz@camk.edu.pl}}

\shortauthors{Zdziarski et al.}

\begin{abstract}
We study the jet in the hard state of the accreting black-hole binary MAXI J1820+070. From the available radio-to-optical spectral and variability data, we put strong constraints on the jet parameters. We find while it is not possible to uniquely determine the jet Lorentz factor from the spectral and variability properties alone, we can estimate the jet opening angle ($\approx1.5\pm1\degr$), the distance at which the jet starts emitting synchrotron radiation ($\sim$3$\times10^{10}$\,cm), and the magnetic field strength there ($\sim$10$^4$\,G), with relatively low uncertainty, as they depend weakly on the bulk Lorentz factor. We find the breaks in the variability power spectra from radio to sub-mm are consistent with variability damping over the time scale equal to the travel time along the jet at any Lorentz factor. This factor can still be constrained by the electron-positron pair production rate within the jet base, which we calculate based on the observed X-ray/soft gamma-ray spectrum, and the jet power, required to be less than the accretion power. The minimum ($\sim$1.5) and maximum ($\sim$4.5) Lorentz factors correspond to the dominance of pairs and ions, and the minimum and maximum jet power, respectively. We estimate the magnetic flux threading the black hole and find the jet can be powered by the Blandford-Znajek mechanism in a magnetically-arrested flow accretion flow. We point out the similarity of our derived formalism to that of core shifts, observed in extragalactic radio sources.
\end{abstract}

\section{Introduction}
\label{intro}

Our knowledge of the structure of extragalactic radio jets is already quite detailed, see, e.g., the review of \citet{Blandford19}. While a number of aspects remains to be determined, e.g., the jet lateral structure \citep{Perlman19}, radio maps provide us the projected structures of the jets, in particular their opening angles. Magnetic fields can be determined via core shifts, which are angular displacements of the position of the radio core between two frequencies (e.g., \citealt{Lobanov98}; \citealt{Zamaninasab14, ZSPT15}, hereafter \core). Superluminal motion allows us to estimate the jet bulk Lorentz factors, $\Gamma$ (e.g., \citealt{Jorstad01, Kellerman04, Lister19}). They can also be independently estimated from radiative models of blazars \citep{Ghisellini15} and from the radio core luminosity function \citep{Yuan18}. The jet power can be estimated from calorimetry of radio lobes \citep{Willott19, Shabala13} and core shifts (e.g., \citealt{Pjanka17}). Finally, the e$^\pm$ pair content can be obtained from comparison of the observed jet powers with theoretical predictions (\citealt{Sikora20} and references therein). 

On the other hand, our knowledge of jets in accreting black-hole (BH) binaries, which are the main class of microquasars, is much more rudimentary. From available radio maps, we can only set upper limits on the jet opening angles (e.g., \citealt{Stirling01, Miller-Jones06}). We can estimate the Lorentz factors of ejected transient blobs, which phenomenon is associated with transitions from the hard spectral state to the soft one, e.g., \citet{Atri20,Wood21}, but this determination strongly depends on the distance to the source. The Lorentz factors of steady compact jets commonly present in the hard spectral state are even more difficult to constrain, with only rough estimates of $\Gamma\gtrsim 1.5$--2 (e.g., \citealt{Stirling01, Casella10, Tetarenko19}). Therefore, an accurate determination of the jet parameters even for a single source would be very important.

Here we study the jet in the transient BH X-ray binary \source during its outburst in 2018. We use the observational data of outstanding quality for that source presented by \citet{Tetarenko21} (hereafter \tet), which gives us an opportunity of such an accurate parameter determination. We interpret these data in terms of the classical model of \citet{BK79} and \citet{Konigl81}. Here flat radio spectra are interpreted in terms of a superposition of synchrotron self-absorbed and optically-thin spectra, spectra above the break frequency are optically-thin synchrotron, and the electron distribution and the magnetic field strength are parametrized by power laws. The jet in the synchrotron-emitting part is assumed to be conical and of a constant bulk-motion velocity. We provide an updated analysis of those data, making corrections to the similar model used in \tet. In particular, \tet followed the formulation of the model that suffered from some errors related to the transformation from the comoving frame to that of the observer. Also, we properly connect the break frequencies in the radio/mm spectra with the propagation time along the jet, and we correct the expressions for jet power. Furthermore, we link the dependencies of energy densities on the distance to the observed hard inverted spectral index, as well as we use additional data from \citet{Rodi21}. This allows us to obtain constraints based on the full radio-through-optical spectrum.

\source was discovered during its outburst in 2018 \citep{Tucker18,Kawamuro18}. The source is relatively nearby, with a distance of $D\approx 2.96\pm 0.33$\,kpc measured based on a radio parallax \citep{Atri20}. Then, \citet{Wood21} determined $D\leq 3.11\pm 0.06$\,kpc based on the proper motion of the moving ejecta during the hard-to-soft state transition. The inclination of the radio jet is $i\approx 64\degr\pm 5\degr$ \citep{Wood21}, while the inclination of the binary is constrained to $i_{\rm b}\approx 66\degr$--$81\degr$ \citep{Torres20}. The BH mass is given by $M\approx (5.95\pm 0.22)\msun/\sin^3 i_{\rm b}$ \citep{Torres20}. 

The data presented in \tet were obtained during a multiwavelength observational campaign from radio to X-rays performed during a 7-h period during the hard state on 2018 April 12 (MJD 58220). We also use the simultaneous IR and optical data obtained by \citet{Rodi21}. During the campaign, the source was in a part of the initial hard state which formed a plateau in the X-ray hardness vs.\ flux diagram \citep{Buisson19}. 

Our theoretical model is presented in Section \ref{jets} and Appendix \ref{general}. In Section \ref{MAXI}, we fit it to the data. In Section \ref{discussion}, we discuss various aspects of our results, and show that our formalism based on time lags between different frequencies of the flat spectrum is equivalent to the formalism based on core shifts. We give our conclusions in Section \ref{conclusions}.

\section{Steady-state jets}
\label{jets}
\subsection{Power-law dependencies}
\label{power_law}

\begin{figure}[t]
\centerline{
\includegraphics[width=7.5cm]{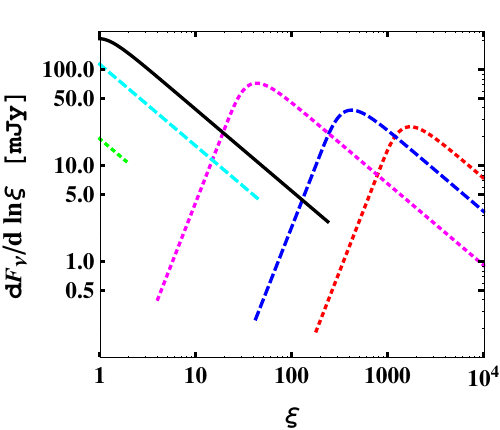}}
  \caption{An example of the spatial structure of the jet emission at different frequencies for $\nu_0=2\times 10^{4}$\,GHz, $F_0=300$\,mJy, $b=1.1$, $a=2 b$, $p=2$. The red dots, blue dashes, magenta dots, cyan dashes, and green dots correspond to $\nu=5.25$, 25.9, 343.5, $1.4\times 10^5$\,GHz, and $5\times 10^{6}$\,GHz, respectively. The black solid curve corresponds to $\nu=\nu_0$. The three lowest and two highest frequency curves end at $\xi_{\rm min}>1$ and at $\xi_{\rm max}$, respectively, beyond which there is no emission at those $\nu$. These values of $\xi_{\rm min}$, $\xi_{\rm max}$ were calculated for $\gamma_{\rm min}=10$, $\nu_{\rm max}=10^7$\,GHz, $i=64\degr$, $\Gamma=3$ and $B_0=10^4$\,G (which correspond to $\gamma_{\rm max}\approx 793$). 
}\label{spatial}
\end{figure}

Following the theoretical interpretation in \tet, we consider a continuous, steady-state, jet in the range of its distance from the BH where it has constant both the bulk Lorentz factor and the opening angle, i.e., it is conical. We consider its synchrotron emission and self-absorption, and assume that the hard, partially self-absorbed, part of the total spectrum results from superposition of spectra from different distances with breaks corresponding to unit self-absorption optical depth \citep{BK79}. We use the formulation of the model of \citet{Konigl81} (which is an extension of the model of \citealt{BK79} for cases with the self-absorbed radio index different from zero) as developed in \citet{Zdziarski19b}, hereafter \zdz. In this model, the jet is assumed to be laterally uniform, which is a good approximation for $i\gg\Theta$, where $\Theta$ is the jet (half) opening angle. We denote the observed and comoving-frame photon frequencies as $\nu$ and $\nu'$, respectively, and introduce the dimensionless distance,
\begin{equation}
\nu'= \frac{\nu (1+z_{\rm r})}{\delta},\quad \delta\equiv \frac{1}{\Gamma(1-\beta\cos i)},\quad \xi\equiv \frac{z}{z_0},
\label{definitions}
\end{equation}
where $z_{\rm r}$ is the cosmological redshift (equal to null in our case), $\Gamma$ and $\beta$ are the jet bulk Lorentz factor and the velocity in units of the light speed, respectively, $z$ is the distance from the BH, and $z_0$ is the distance at which the jet becomes optically thin to self-absorption at all considered frequencies. As in \citet{Konigl81}, we assume the electron differential density distribution, $n(\gamma,\xi)$, and the magnetic field strength, $B$, are parameterized by power-law dependencies,
\begin{equation}
R(\xi)= z_0\xi\tan\Theta,\, n(\gamma,\xi)=n_0 \xi^{-a}\gamma^{-p},\, B(\xi)=B_0 \xi^{-b},
\label{pl}
\end{equation}
where $R$ is the jet radius and $\gamma$ is the Lorentz factor of the emitting electrons in the jet comoving frame, with $\gamma_{\rm min}\leq \gamma\leq \gamma_{\rm max}$. The quantities $R$ and $z$ are measured in the local observer's frame, while $n$ and $B$ are given in the comoving frame (for notational simplicity, we skip the primes). For a conserved electron number along the jet and conserved magnetic energy flux dominated by toroidal fields, we have $a=2$ and $b=1$, corresponding to the spectral index of $\alpha=0$, independent of the value of $p$ \citep{BK79}. Here, we define $\alpha$ by the energy flux density of $F_\nu \propto \nu^\alpha$. If either the electron or magnetic energy is dissipated, $a>2$, $b>1$, respectively. Then, the emission weakens with the distance and the synchrotron spectrum in the partially self-absorbed frequency range becomes harder than in the conserved case, $\alpha>0$. The spectral indices of partially self-absorbed and optically thin synchrotron emission are
\begin{equation}
\alpha=\frac{5 a+3 b+2(b-1)p-13}{2a-2+b(p+2)},\quad \alpha_{\rm thin}=\frac{1-p}{2},
\label{alpha}
\end{equation}
respectively (\citealt{Konigl81}; \zdz). Using a delta-function approximation to the single-electron synchrotron spectrum at $\gamma^2\gg 1$ (assumed hereafter), the synchrotron frequency for a given $\gamma$ and $\xi$, and its range emitted by the jet are
\begin{align}
\frac{h\nu'}{m_{\rm e}c^2}= &\frac{B_0\xi^{-b}}{B_{\rm cr}}\gamma^2,\label{syn_nu}\\
\frac{h\nu'_{\rm min}}{m_{\rm e}c^2}= \frac{B_0 \xi_{\rm M}^{-b}}{B_{\rm cr}}\gamma_{\rm min}^2&, \quad
\frac{h\nu'_{\rm max}}{m_{\rm e}c^2}= \frac{B_0}{B_{\rm cr}}\gamma_{\rm max}^2, 
\label{nu_range}
\end{align}
respectively. Here $z_0\xi_{\rm M}$ is the distance at which the jet terminates, $h$ is the Planck constant, $B_{\rm cr}={2\pi m_{\rm e}^2 c^3/(e h)}\approx 4.414\times 10^{13}$\,G is the critical magnetic field strength, and $m_{\rm e}$ and $e$ is the electron mass and charge, respectively. The spectral density of the synchrotron emission for a single jet parameterized by Equation (\ref{pl}) and for $\nu_{\rm min}\leq\nu\leq \nu_{\rm max}$ is then given by (see equation A5 of \zdz),
\begin{align}
F_\nu &\simeq  
F_0 \left(\frac{\nu}{\nu_0}\right)^{\frac{5}{2}} \int_{\xi_{\rm min}}^{\xi_{\rm max}} {\rm d}\xi\,  \xi^{1+b/2}\left\{1-\exp[-\tau(\frac{\nu}{\nu_0},\xi)]\right\}, \label{integral}\\
F_0&\equiv{(1+z_{\rm r})^\frac{7}{2}(m_{\rm e}h \delta)^\frac{1}{2} \pi C_1(p) z_0^2 \nu_0^\frac{5}{2} \tan\Theta\sin i\over 6 c C_2(p) (B_0/ B_{\rm cr})^\frac{1}{2} D^2}.
\label{F0}
\end{align}
Here $F_0$ is a constant proportional to the bolometric flux, $\tau(\nu/\nu_0,\xi)$ is the synchrotron self-absorption optical depth, $\nu_0$ is the break frequency, see Equation (\ref{tau}) below, and $C_1(p)$, $C_2(p)$ follow from averaging the synchrotron emission and absorption coefficients over the pitch angle,
\begin{align}
&C_1(p) = {3^{p+4\over 2} \Gamma_{\rm E}\left(3p-1\over 12\right) \Gamma_{\rm E}\left(3p+19\over 12\right) \Gamma_{\rm E}\left(p+1\over 4\right) \over 2^5\pi^{1\over 2}\Gamma\left(p+7\over 4\right)},
\label{c1}\\
&C_2(p)={3^{p+3\over 2} \Gamma_{\rm E}\left(3p+2\over 12\right) \Gamma_{\rm E}\left(3p+22\over 12\right) \Gamma_{\rm E}\left(p+6\over 4\right)\over 2^4\pi^{\frac{1}{2}} \Gamma_{\rm E}\left(p+8\over 4\right)}\label{c2}
\end{align}
(cf.\ \citealt{Jones74, ZLS12}), where $\Gamma_{\rm E}$ is the Euler Gamma function. In the extragalactic case, $D$ is the luminosity distance. The lower and upper limits of the integral (\ref{integral}) are,
\begin{align}
&\xi_{\rm min}(\nu)=\max\left[1,\left(\frac{B_0 m_{\rm e}c^2\gamma_{\rm min}^2}{h\nu' B_{\rm cr}}\right)^{\frac{1}{b}}\right],\label{ximin}\\
&\xi_{\rm max}(\nu)=\min\left[\left(\frac{\nu_{\rm max}}{\nu}\right)^{\frac{1}{b}},\xi_{\rm M}\right],
\label{ximax}
\end{align}
respectively. Figure \ref{spatial} shows an example of the spatial dependencies of the emission along the jet at different frequencies for $\gamma_{\rm min}=10$ and $\nu_{\rm max}=10^7$\,GHz. For $\gamma_{\rm min}\gtrsim 30$, the emission at all frequencies in this case would be in the optically-thin regime only, cf.\ Equation (\ref{gamma_nu}) below. We note that above we have assumed the single-electron synchrotron emission is isotropic in the plasma frame, which is strictly valid for a tangled magnetic field.

\begin{figure}[t]
\centerline{
\includegraphics[width=7.cm]{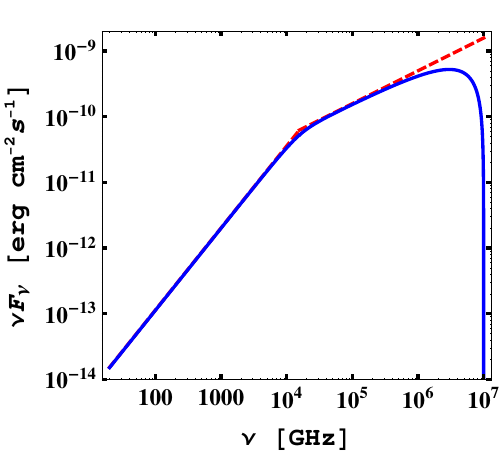}}
  \caption{An example of the jet synchrotron spectrum for $\nu_0=2\times 10^{4}$\,GHz, $F_0=300$\,mJy, $b=1.1$, $a=2 b$, $p=2$, $\nu_{\rm max}=10^7$\,GHz, $i=64\degr$. This spectrum is virtually independent of $\xi_{\rm min}(\nu)$ in the shown range of $\nu$, as long as $\xi_{\rm min}(\nu)\ll \xi_\nu$. The blue curve shows the accurate spectrum of Equation (\ref{integral}), and the red dashes show the approximation of Equation (\ref{solution}). The gradual high-energy cutoff of the accurate spectrum is due to $\xi_{\rm max}$ decreasing with increasing $\nu$ and reaching unity for $\nu_{\rm max}$. 
}\label{model_spectra}
\end{figure}

Then, power-law dependencies assuming $\xi_{\rm min}=1$, $\xi_{\rm max}= \infty$ in the optically-thick and optically thin cases are (cf.\ equation A11 in \zdz)
\begin{equation}
F_\nu \simeq 2 F_0
\begin{cases} \displaystyle{\Gamma_{\rm E}\left[\frac{2 a-6+b(p+1)}{2 a-2+b(p+2)}\right]}\frac{(\nu/\nu_0)^{\alpha}}{4+b}
 , &\nu\ll \nu_0;\cr
 \displaystyle{\frac{(\nu/\nu_0)^{\alpha_{\rm thin}}}{2 a-6+b(p+1)}}, &\!\!\!\!\!\!\!\!\!\!\!\!\!\!\!\!\nu_0\ll \nu\ll \nu_{\rm max}.\cr\end{cases}
\label{solution}
\end{equation}
Figure \ref{model_spectra} shows an example comparison of the accurate spectrum of Equation (\ref{integral}) with these power-law approximations. We see they are inaccurate around $\nu_0$ as well as close to $\nu_{\rm max}$, where they fail to reproduce the gradual cutoff of the accurate spectrum. While the power-law asymptotic solutions intersect at a $\nu$ slightly different from $\nu_0$, that frequency has no physical meaning since the actual spectrum in that range does not follow the broken-power law form, see Figure \ref{model_spectra}. We can define a broken-power law approximation by taking the minimum of the two branches in Equation (\ref{solution}).

The optical depth along a line of sight crossing the jet spine can be written as
\begin{equation}
\tau(\nu/\nu_0,\xi)=(\nu/\nu_0)^{-(p+4)/2} \xi^{1-a-b(p+2)/2},
\label{tau}
\end{equation}
where $\nu_0$ is defined by $\tau(\nu=\nu_0,\xi=1)=1$. The place $\xi=1$, or $z=z_0$, corresponds to the jet being optically thin for all $\nu\geq \nu_0$. There is no synchrotron emission\footnote{We note that since the partially optically-thick emission of a jet at $z>z_0$ would remain almost unaffected if there were still emission following the scaling of Equation (\ref{pl}) at $z<z_0$ (which would, however, decrease the actual value of $z_0$ and increase $\nu_0$), it is also possible to formulate the structure of the partially optically-thick part without invoking $z_0$ and $\nu_0$. Such a formulation is presented in Equations (\ref{B_0eq}--\ref{B_0f}) in Appendix \ref{general}.} at $z<z_0$, and thus $z_0$ corresponds to the onset of the jet emission. The relationship of $\nu_0$ to the jet parameters is given by equation (A8) of \zdz. We express it here as a formula for the normalization of the electron distribution,
\begin{equation}
n_0=\left(\frac{B_{\rm cr}}{B_0\delta}\right)^{1+\frac{p}{2}}\!\! \left[\frac{h \nu_0(1+z_{\rm r})}{m_{\rm e}c^2}\right]^{2+\frac{p}{2}}\!\! \frac{\alpha_{\rm f} \sin i}{C_2(p) \pi \sigma_{\rm T} z_0 \tan\Theta},
\label{tau1}
\end{equation}
where $\alpha_{\rm f}$ is the fine-structure constant and $\sigma_{\rm T}$ is the Thomson cross section. From Equation (\ref{tau}), the distance along the jet at which $\tau(\nu,\xi_\nu)=1$ at $\nu \lesssim\nu_0$ is
\begin{equation}
\xi_\nu=\left(\frac{\nu}{\nu_0}\right)^{-q}\!\!,\,\, q\equiv\frac{p+4}{2 a+b p+2b-2},\,\, z_\nu=z_0\xi_\nu.
\label{xi_nu}
\end{equation}
For $a=2$ and $b=1$, we have $q=1$ at any $p$. This distance is very close to that at which most of the flux at a given $\nu$ is emitted, which can be defined by the maximum of ${\rm d}F_\nu(\xi)/{\rm d}\,\ln\xi$, see Figure \ref{spatial}, and can be calculated using Equation (\ref{integral}). For example, at $a=2.2$, $b=1.1$, and $p=2$, that maximum is at $\xi\approx 1.19\xi_\nu$. The emission around the peak has a broad spatial distribution; the 50\% values of the maximum flux are reached at $\xi=0.65\xi_\nu$ and $3.64\xi_\nu$.

Then, the Lorentz factor responsible for the bulk of emission at $\xi_\nu$ is
\begin{equation}
\gamma_\nu=\left(\frac{B_{\rm cr}}{B_0}\frac{h\nu_0}{\delta m_{\rm e}c^2}\right)^{1/2} \left(\frac{\nu}{\nu_0}\right)^{(1-b q)/2},
\label{gamma_nu}
\end{equation}
which is usually weakly dependent on $\nu$. While the integral spectrum of Equation (\ref{integral}) is valid for any $\gamma_{\rm min}$, the asymptotic power-laws of Equation (\ref{solution}) require $\gamma_{\rm min}$ to be by a factor of at least a few lower than $\gamma_\nu$ for values of $\nu$ of interest (in the range $<\nu_0$) and $\gamma_{\rm max}$ is required to be a factor of a few larger than $\gamma_\nu$. If a high-energy cutoff is observed, an additional constraint follows from it, see Equation (\ref{syn_nu}).

If we know $\alpha$ and $\alpha_{\rm thin}$, we still cannot determine the values of $a$ and $b$ separately. However, a likely possibility is that the ratio between the electron and magnetic energy densities remains constant, i.e., maintaining the same degree of equipartition along the jet, in which case $a=2 b$. We define an equipartition parameter as the ratio of the energy densities, 
\begin{equation}
\beta_{\rm eq}\equiv {u_{\rm p}\over B^2/8\pi}={n_0 m_{\rm e} c^2 (1+k_{\rm i})(f_E- f_N)\over B_0^2/8\pi},\label{betaeq}
\end{equation}
where
\begin{equation}
f_E\equiv \begin{cases} {\gamma_{\rm max}^{2-p}-\gamma_{\rm min}^{2-p}\over 2-p}, &p\neq 2;\cr
\ln {\gamma_{\rm max}\over \gamma_{\rm min}},& p=2,\cr
\end{cases}\quad
f_N\equiv \frac{\gamma_{\rm min}^{1-p}-\gamma_{\rm max}^{1-p}}{p-1},
\label{fe_fn}
\end{equation}
the second equality in Equation (\ref{betaeq}) is at $z_0$, $u_{\rm p}$ is the particle energy density, $k_{\rm i}$ accounts for the energy density in particles other than the power-law electrons, in particular in ions (excluding the rest energy), and $p>1$ has been assumed in the expression for $f_N$. For $a=2b$, $\beta_{\rm eq}$ is constant along the jet (provided $k_{\rm i}$ is also constant) at $z\geq z_0$, which yields
\begin{equation}
\alpha=\frac{(b-1)(13+2 p)}{b(p+6)-2},\quad q=\frac{p+4}{b(p+6)-2}.
\label{alpha_q}
\end{equation}
Below, we use $\beta_{\rm eq}$ and $a=2b$ to constrain the jet parameters. 

We note that the case of $a>2$ requires that either $\gamma_{\rm min}$ decreases or the electrons removed from their power-law distribution move to some low energies below $\gamma_{\rm min}$ (with negligible emission). Since we assume $\gamma_{\rm min}$ to be constant along the jet, the latter has to be the case.

We next consider the difference between the arrival times of two photons. The first photon, at $\nu_1$, is emitted toward the observer at $\xi_{\nu_1}$. The second photon, with $\nu_2<\nu_1$, is emitted at $\xi_{\nu_2}$ by the same comoving point of the jet after the time $\Delta t_{\rm e}$, which is further downstream in the jet by $\beta c \Delta t_{\rm e}$. Since the jet moves at an angle $i$ with respect to the line of sight, the distance of the emitting point to the observer will become shorter during this time by $\beta c \Delta t_{\rm e}\cos i$. For an observed difference in the arrival times of $\Delta t_{\rm a}$, the intrinsic separation between the emission points (measured in the local observer's frame) will be
\begin{equation}
z_{\nu_2}-z_{\nu_1}=z_0(\xi_{\nu_2}-\xi_{\nu_1})=\frac{\Delta t_{\rm a}\beta c}{(1-\beta \cos i)(1+z_{\rm r})}.
\label{Delta_t}
\end{equation}
At frequencies $\leq \nu_0$, $\xi_\nu$ follows from Equation (\ref{xi_nu}). Here, we have also taken into account the redshift, making this expression correct also for an extragalactic source. Given an observed $\Delta t_{\rm a}$, Equations (\ref{xi_nu}) and (\ref{Delta_t}) imply
\begin{align}
&z_0=\frac{\Delta t_{\rm a}\nu_0^{-q}\beta c}{(1-\beta \cos i)\left(\nu_2^{-q}-\nu_1^{-q}\right)(1+z_{\rm r})}=\frac{t_0 c \beta\Gamma\delta}{1+z_{\rm r}},
\label{z0}\\
&t_0\equiv \frac{\Delta t_{\rm a}}{\Delta \xi}=\frac{\Delta t_{\rm a}\nu_0^{-q}}{\nu_2^{-q}-\nu_1^{-q}},
\label{t0}
\end{align}
where $t_0$ is the implied lag time between the BH and $z_0$, which can be obtained from time lag data if $\nu_0$ and $q$ are known (from the spectrum).

Appendix \ref{general} provides general solutions for $z_\nu$, $B$, $\Theta$ and $n$ to the equations in this section assuming the validity of Equation (\ref{solution}) for $\nu<\nu_0$ as functions of $b$ (assuming $a=2 b$), $p$, $\nu_0$, $F_0$, $t_0$, $i$ and $D$, as well as $\beta_{\rm eq}$, $\gamma_{\rm min}$ and $\gamma_{\rm max}$. 

\subsection{The jet power}
\label{Power}

The jet power can be calculated using standard expressions. Note that it is defined in terms of the proper enthalpy rather than the energy density, e.g., \citet{Levinson06}, \citet{Zdziarski14b}. Then, the component of the jet power due to both the relativistic electrons and magnetic fields (assuming they are predominantly toroidal at $z_0$) including both the jet and counterjet for $a=2b$ and $k_{\rm i}=0$ at $z\geq z_0$ is 
\begin{equation}
P_B+P_{\rm e}=\left(\frac{1}{2}+\frac{\beta_{\rm eq}}{3}\right)\! c\beta(B_0 z_0 \Gamma \tan\Theta)^2 \xi^{2-2 b}.
\label{Prel}
\end{equation}
The usable power in cold ions at any $z$ (calculated at $z_0$), is 
\begin{equation}
P_{\rm i}=2\pi \mu_{\rm e}n_0 f_N\!\left(\!1-\frac{2n_+}{n_{\rm e}}\!\right)\! m_{\rm p}c^3 \beta\Gamma(\Gamma-1)(z_0\tan\Theta)^2,\label{Pcold}
\end{equation}
where $n_{\rm e}$ and $n_+$ is the density of both electrons and positrons (whose ratio is assumed to be constant along the jet), and positrons only, respectively, $\mu_{\rm e}=2/(1+X)$ is the electron mean molecular weight, $X$ ($\approx 0.7$ for the cosmic composition) is the H mass fraction, $m_{\rm p}$ is the proton mass, and $f_N$ is given by Equation (\ref{fe_fn}). This is the power in ions that has to be supplied to the jet, and then it can be dissipated, hence the factor $(\Gamma-1)$. Equation (\ref{Pcold}) neglects the possible presence of background electrons being piled up at $\gamma<\gamma_{\rm min}$ already at $z_0$. On the other hand, $a>2$ at a constant $\gamma_{\rm min}$ requires that leptons removed from $n(\gamma,\xi)$ at $z>z_0$ do appear at some low energies below $\gamma_{\rm min}$ (with negligible emission).

We note that in a steady state, $2 n_+/n_{\rm e}=2 \dot N_+/\dot N_{\rm e}$, where $2\dot N_+$ is the total rate of advection upstream of e$^\pm$ pairs produced at the jet base, and $\dot N_{\rm e}$ is the total lepton flow rate,
\begin{equation}
\dot N_{\rm e} \approx 2\pi n_0 f_N c\beta\Gamma (z0 \tan\Theta)^2.
\label{ndot_e}
\end{equation}
This implies
\begin{equation}
P_{\rm i}=\mu_{\rm e} m_{\rm p}c^2 (\Gamma-1)\left(\dot N_{\rm e}- 2\dot N_+\right)\geq 0.
\label{Pcold3}
\end{equation}
Since the pair production rate at the jet base, $\dot N_+$ (Section \ref{pairs}), is independent of $\Gamma$, the condition of $P_{\rm i}\geq 0$ can give a lower limit on $\Gamma$.

In any jet model, the total usable jet power is approximately constrained by the accretion power, 
\begin{equation}
P_{\rm j}=P_B+P_{\rm e}+ P_{\rm i}\lesssim \dot M c^2= \frac{L}{\epsilon_{\rm eff}}
\label{Pjet_dotM}
\end{equation}
where $\dot M$ is the mass accretion rate, $L$ is the bolometric luminosity and $\epsilon_{\rm eff}\sim 0.1$ is the accretion efficiency. This then gives an upper limit on $\Gamma$. The limit $\dot M c^2$ can be exceeded if the rotation of the BH is tapped, but only by a factor $\lesssim$1.3 and for a maximally rotating BH, see the references in Section \ref{BZ} below.

Finally, we consider the power lost in synchrotron emission. It equals the synchrotron luminosity emitted by both jets in all directions (which is Lorentz invariant). Since $\nu_0$ depends on the direction and the partially self-absorbed emission is not isotropic in the comoving frame, we neglect its effect and assume the entire emission is optically thin and isotropic in that frame, which is a good approximation for hard electron distributions with $p\lesssim 2.5$ or so. This gives
\begin{align} 
P_{\rm S}&\approx \frac{1}{3}(B_0\tan\Theta)^2\sigma_{\rm T} c z_0^3 n_0\Gamma f_{E2} f_\xi,\nonumber\\
f_{E2}&\equiv \begin{cases} {\gamma_{\rm max}^{3-p}-\gamma_{\rm min}^{3-p}\over 3-p}, &p\neq 3;\cr
\ln \frac{\gamma_{\rm max}}{\gamma_{\rm min}},& p=3,\cr
\end{cases}\label{P_S}\\
f_\xi&\equiv \int_1^{\infty}{\rm d}\xi \xi^{2-2 b- a}=\frac{1}{2 b+a-3},
\nonumber
\end{align}
where $2 b+a>3$ is assumed. This $P_{\rm S}$ approximately equals the intrinsic luminosity of both jets,
\begin{equation}
L_{\rm jet}\approx 8\pi D^2 \delta^{-3}\Gamma \int_{0}^{\nu_{\rm max}} {\rm d}\nu F_\nu,
\label{Ljet}
\end{equation}
where $F_\nu$ is for the approaching jet, $\nu_{\rm max}$ is given by Equation (\ref{syn_nu}) and the transformation law is for a stationary jet emitting isotropically in its comoving frame \citep{Sikora97}. For self-consistency of our equations, $P_{\rm S}\ll P_{\rm j}$ is required. 

\subsection{Pair production}
\label{pairs}

As we see from Equation (\ref{Pcold3}), the jet power in ions (given an observed synchrotron spectrum) strongly depends on the abundance of e$^\pm$ pairs. In the case of extragalactic jets, there are strong indications that they dominate by number, though most of the rest mass is usually still in ions \citep{Sikora20}. In the case of jets in microquasars, this is uncertain. An important issue for that is the origin of pairs. A likely mechanism is pair production in photon-photon collisions by photons produced close to the BH.
 
\begin{figure}[t]
\centerline{
\includegraphics[width=8cm]{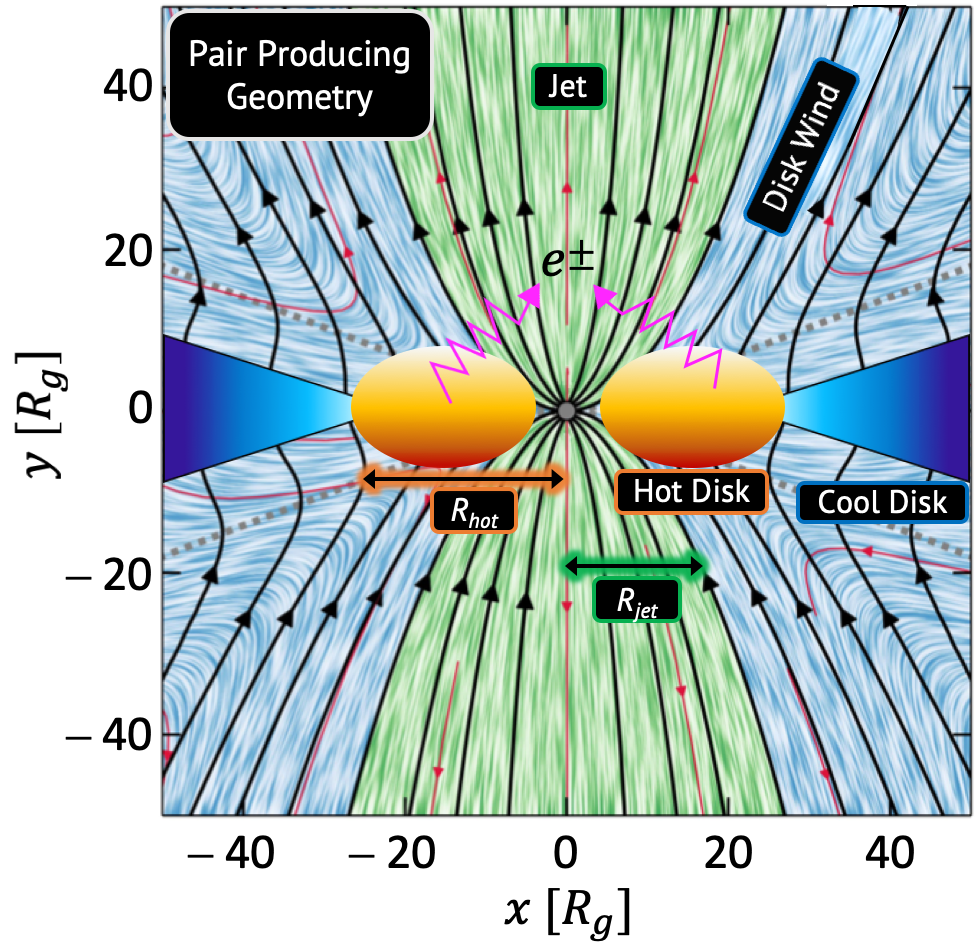}}
  \caption{A sketch of of the pair-producing geometry based on fig.\ 3.9 of \citet{Tchekhovskoy15}, which shows the result of his 3D GRMHD simulation for magnetically-arrested accretion on a BH with the spin parameter of $a_*=0.99$. In our case the disk is hot in its inner part (up to the radius $R_{\rm hot}$) and surrounded by a cool outer disk. We consider e$^\pm$ pair production within the jet base (shown in green), which is devoid of matter, with the wavy arrows representing pair-producing photons. We denote the characteristic jet radius of the pair-producing region as $R_{\rm jet}$. In addition, pairs are produced within the hot disk, but it is magnetically shielded from the jet base. The black solid curves show the poloidal magnetic field. 
}\label{jet_pairs}
\end{figure}

Pairs can be produced within the hot flow, e.g, \citet{Svensson87}. Since the Larmor radius of either a proton or an electron is orders of magnitude lower than $R_{\rm g}$ (where $R_{\rm g}=G M/c^2$ is the gravitational radius), the magnetic base of the jet is shielded from the hot plasma, and pairs produced in the accretion flow cannot enter the jet. On the other hand, pairs can also be produced within the magnetic base of the jet, outside the hot plasma \citep{Sikora20}. There, photon-photon collisions will create e$^\pm$ pairs in an environment devoid of background matter, thus strongly reducing the rate of pair annihilation \citep{B99_pairs}. A possible geometry (see also \citealt{Henri91,Ferreira06}) is shown in Figure \ref{jet_pairs}. From an observed hard X-ray spectrum and a radius, $R_{\rm hot}$, of the emitting hot plasma (inferred, e.g., from X-ray spectroscopy; \citealt{Bambi20}), we can estimate the average photon density within the jet base, which then gives us the rate of pair production per unit volume, $\propto R_{\rm hot}^{-4}$. We approximate the pair-producing volume as two cylinders with the height $R_{\rm hot}$ and the characteristic radius of the jet, $R_{\rm jet}$, i.e., $V=2\pi R_{\rm jet}^2 R_{\rm hot}$. We can then write (assuming $R_{\rm jet}\leq R_{\rm hot}$) the total lepton production rate as
\begin{equation}
2\dot N_+ = A_{\gamma\gamma} R_{\rm hot}^{-3} R_{\rm jet}^2,
\label{Nplus0}
\end{equation}
where the factor $A_{\gamma\gamma}$ would follow from detailed calculations. Depending on the equilibrium density of the pairs, some of them would annihilate, and some would be advected to the BH, reducing the effective $2 \dot N_+$. We address this issue for the case of \source in Section \ref{analytical}. 
 
\subsection{The Blandford-Znajek mechanism}
\label{BZ}

We can also estimate the jet power in the framework of the model with extraction of the rotational power of the BH \citep{BZ77}. The jet power in this case depends on the magnetic flux, $\Phi_{\rm BH}$, threading the BH (on one hemisphere), which can be written as
\begin{equation}
\Phi_{\rm BH}=\phi_{\rm BH}(\dot M c)^{1/2} R_{\rm g},
\label{phi_BH}
\end{equation}
where $\phi_{\rm BH}$ is a dimensionless magnetic flux. Its maximum value of $\approx$50 is obtained in magnetically arrested disks (MAD; \citealt{Narayan03,BK74}), as it was found in GRMHD simulations of MAD accretion (\citealt{Tchekhovskoy11, McKinney12}; see its more accurate value in \citealt{Davis20}). Then it has been found that
\begin{equation}
P_{\rm j}\approx 1.3\left(\frac{\phi_{\rm BH}}{50}\right)^2 h_{0.3} a_*^2\dot M c^2,
\label{P_tot}
\end{equation}
where $a_*$ is the BH spin parameter and $h_{0.3}$ is defined by the half-thickness of the disk being $H_{\rm disk}=R_{\rm disk} 0.3 h_{\rm 0.3}$ \citep{Davis20}. This maximum differs from that of Equation (\ref{Pjet_dotM}) by the factor $1.3 h_{0.3} a_*^2$. In the spectrally hard state, the disk is most likely hot, in which case $h_{0.3}\sim 1$.

\begin{figure*}
\centerline{
\includegraphics[width=1.05\textwidth]{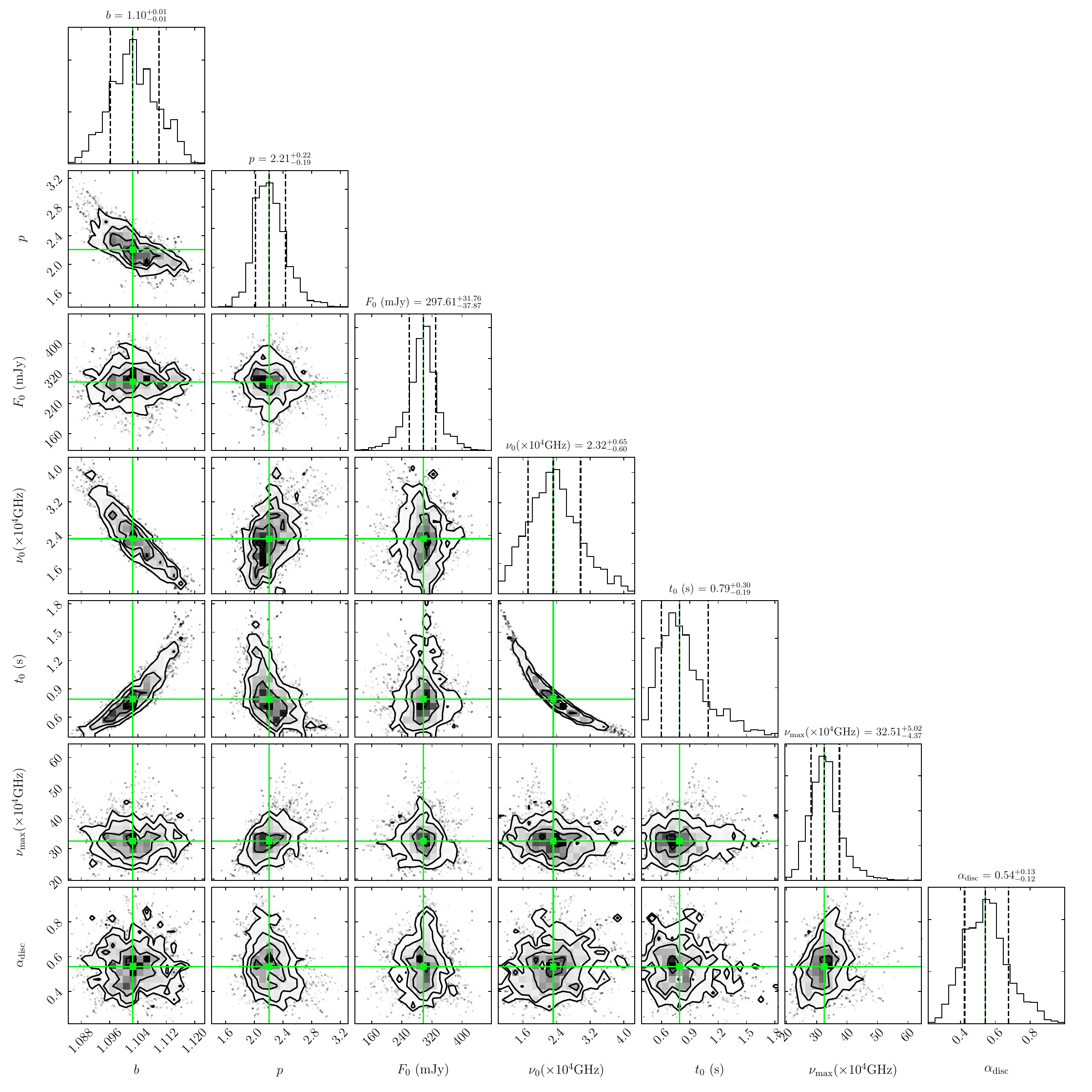}}
  \caption{MCMC fit results for the seven model-independent quantities, which require only the assumptions of $a=2 b$, see Section \ref{analytical}. Here and in Figure \ref{fits_full} below, the panels show the histograms of the one-dimensional posterior distributions for the model parameters, and the two-parameter correlations, with the best-fitting values of the parameters indicated by green lines/squares. The best-fit results for fitted quantities are taken as the medians of the resulting posterior distributions, and are shown by the middle vertical dashed lines in the distribution panels. The surrounding vertical dashed lines correspond approximately to a $1\sigma$ uncertainty. 
}\label{basic_fits}
\end{figure*}

We can estimate $\phi_{\rm BH}$ using the magnetic field strength measured far from the BH by using the conservation of the magnetic flux. Specifically, we use the expected equality between the poloidal and toroidal field components at the Alfv{\'e}n surface (in the observer's frame), which radius, for strongly magnetized jets, approaches the light cylinder radius, $R_{\rm LC}$ \citep{Lyubarsky10}. This implies $\Gamma \langle B'_\phi\rangle\approx (R/R_{\rm LC}) B_{\rm p}$, where $B_{\rm p}$ is the poloidal field (which has the same value in the comoving and BH frames) and $\langle B'_\phi\rangle$ is the average toroidal field strength in the comoving frame, denoted by $B$ in the remainder of this paper. Then, the toroidal field dominates at $z$ satisfying $R(z)\gg \Gamma R_{\rm LC}$, and, presumably, at $z\geq z_0$. The magnetic flux using this method was determined for a sample of radio loud active galactic nuclei in \citet{Zamaninasab14} and \core. We use the resulting formula as derived in \core,
\begin{equation}
\Phi_{\rm j}={2^{3/2}\pi R_{\rm H}s z_0 B_0(1+\sigma)^{1/2}\over \ell a_*},
\label{phi_jet}
\end{equation}
which allows us to estimate $\phi_{\rm BH}$ for a given $a_*$ by setting $\Phi_{\rm j}=\Phi_{\rm BH}$. Here $R_{\rm H}=[1+(1-a_*^2)^{1/2}] R_{\rm g}$ is the BH horizon radius, $\ell\la 0.5$ is the ratio of the field and BH angular frequencies, and $s$ is the scaling factor relation between the jet opening angle and the bulk Lorentz factor \citep{Komissarov09, Tchekhovskoy09}, limited by causality to $\lesssim 1$, 
\begin{equation}
\Theta\approx s \sigma^{1/2}/\Gamma.
\label{opening}
\end{equation}
Here, $\sigma$ is the magnetization parameter, which is defined as the ratio of the proper magnetic enthalpy to that for particles including the rest energy,
\begin{equation}
\sigma \equiv {B^2/4\pi\over \eta u_{\rm p}+\rho c^2}=\frac{2}{\beta_{\rm eq}}\left[\eta+\frac{\mu_{\rm e}m_{\rm p} (1 -2 n_+/n_{\rm e})f_N}{m_{\rm e} (f_E-f_N)(1+k_{\rm i})}\right]^{-1},
\label{sigma}
\end{equation}
where $\rho$ is the rest-mass density, and $4/3<\eta<5/3$ is the particle adiabatic index. The second equality relates $\sigma$ to $\beta_{\rm eq}$ assuming that the only ions are those associated with the power law electrons (i.e., neglecting the possible presence of ions associated with electrons with $\gamma<\gamma_{\rm min}$, e.g., with a quasi-Maxwellian distribution). For $p>2$ and a large enough $\gamma_{\rm max}$, $f_E/f_N\approx \gamma_{\rm min}(p-1)/(p-2)$. Then, for $\beta_{\rm eq}(1 -2 n_+/n_{\rm e})/\gamma_{\rm min}\gg m_{\rm e}/m_{\rm p}$, we have $\sigma\ll 1$. 

\section{Application to \source}
\label{MAXI}

Here, we apply the model of Section \ref{jets} to the source. We use the VLA fluxes at 5.25, 7.45, 8.5, 11.0, 20.9, 25.9 and the ALMA flux at 343.5 (from table 1 in \tet). We also use the IR flux at $1.4\times 10^5$ from VLT/HAWK-I, and the optical flux at $3.9\times 10^5$\,GHz from NTT/ULTRACAM (\tet), and the 13 fluxes between 1.37 and $7.00\times 10^5$\,GHz from the VLT X-shooter and the \integral/OMC flux at $5.66\times 10^5$\,GHz \citep{Rodi21}. All of the IR/optical fluxes have been de-reddened with $E(B\!-\!V)=0.18$ \citep{Tucker18}, as assumed in \tet. We use the time lags between 25.9\,GHz and lower frequencies, 11.0\,GHz and lower frequencies, and between 343.5\,GHz and lower frequencies. The lags are given in tables 3 and 4 of \tet. This gives us 23 spectral measurements and 14 time lags. We present analytical and numerical estimates in Sections \ref{analytical} and \ref{mcmc}, respectively. We assume the ratio between the electron and magnetic energy densities to be constant along the jet, i.e., $a=2 b$. 

In our fits below, we use the Markov-Chain Monte Carlo (hereafter MCMC) technique with wide uniform priors, see \tet for details. We assume $i=64\degr\pm 5\degr$ \citep{Wood21} with a Gaussian prior, and $D=2.96\pm 0.33$\,kpc \citep{Atri20} with a Gaussian prior, but truncated at the upper limit of $D_{\rm max}=3.11$\,kpc found by \citet{Wood21}.

We assume the observed spectrum is from the approaching jet only. At $i\approx 64\degr$, this assumption is satisfied only roughly. The ratio of the jet-to-counterjet fluxes in the optically-thick part of the spectrum is given by $[(1+\beta\cos i)/(1-\beta\cos i)]^{(7+3 p)/(4+p)}$ (which follows from \citealt{ZLS12}), which is $\approx$7 at the fitted $p\approx 2.21$ (see below).  

\subsection{The initial fit and analytical estimates}
\label{analytical}

\begin{table*}[t!]
\centering
\caption{The basic parameters of the jet in \source. }
\begin{tabular}{cccccccc}
\hline
$b$ & $p$ & $\nu_0$ & $F_0$ & $\nu_{\rm max}$ & $\alpha_{\rm disk}$ & $F_{\rm disk}$ & $t_0$\\
 &  & $10^4$\,GHz & mJy & $10^4$\,GHz &  & mJy & s\\
\hline
$1.10^{+0.01}_{-0.01}$ & $2.21^{+0.22}_{-0.19}$ & $2.32^{+0.65}_{-0.60}$ & $298^{+31}_{-38}$ & $32.5^{+5.0}_{-4.4}$ & $0.54^{+0.13}_{-0.12}$ & $30.4_{-5.9}^{+6.0}$ & $0.79^{+0.30}_{-0.19}$\\
\hline
\end{tabular}
\tablecomments{The fits are done with the MCMC method assuming $a=2 b$. $F_{\rm disk}$ gives the flux density at $10^5$\,GHz. 1 mJy $\equiv 10^{-26}$\,erg\,cm$^{-2}$\,s$^{-1}$\,Hz$^{-1}$.}
\label{basic}
\end{table*}

\begin{figure}
\centerline{
\includegraphics[width=\columnwidth]{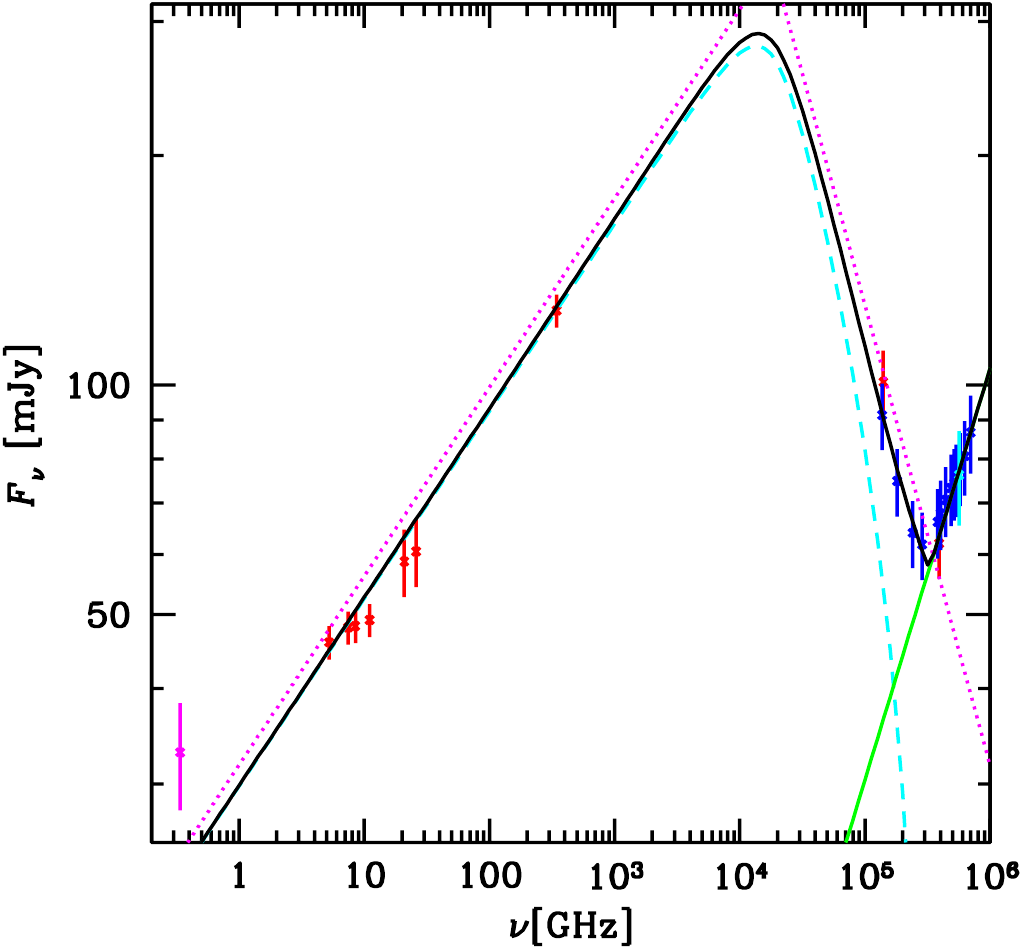}}
  \caption{The radio-to-optical spectrum from \tet (VLA, ALMA, VLT/HAWK-I, NTT/ULTRACAM; red error bars), the 339\,MHz measurement (VLITE, magenta error bar; \citealt{Polisensky18}), and from the VLT/X-shooter (blue error bars) and the \integral/OMC (cyan error bar) as obtained by \citet{Rodi21}, but with the de-reddening correction for $E(B\!-\!V)=0.18$ \citep{Tucker18}. The error bars for the radio and sub-mm measurements of \tet are the square roots of the squares of their statistical and systematic errors, and 10\% systematic errors are assumed for the IR and optical measurements. The spectrum above 5 GHz is fitted by the jet model of Equation (\ref{integral}) using the best-fit parameters shown in Figure \ref{basic_fits} (cyan dashed curve) and a phenomenological power law approximating the disk component with $\alpha_{\rm disk}=0.53$ (green solid curve). The former is virtually independent of $\xi_{\rm min}$ within the ranges obtained in the full fits (Equation \ref{ximin}; Section \ref{mcmc}), so it can be assumed to be unity. The sum is shown by the solid black curve. The corresponding asymptotic optically thick and optically thin spectra of Equation (\ref{solution}) are shown by the magenta dotted lines.}
\label{spec}
\end{figure}

\begin{figure}
\centerline{
\includegraphics[width=7cm]{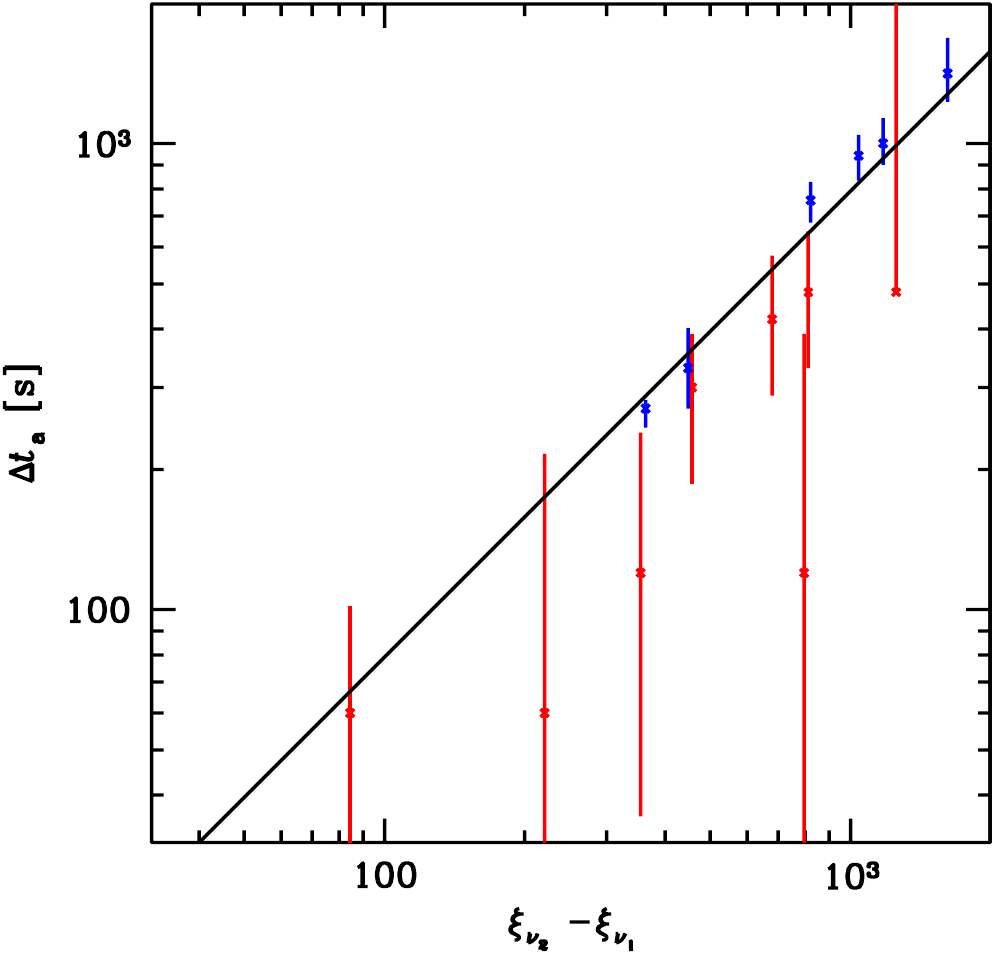}}
  \caption{The time lags measured by \tet vs. the theoretically expected distance in units of $z_0$ between the emission points for the partially-self-absorbed part of the spectrum. The blue and red symbols correspond to the lags between 343.5\,GHz and radio frequencies (5.25--25.9\,GHz), and within the radio frequencies, respectively, where we assumed a constant ratio between the electron and magnetic energy densities, $\xi_\nu= (\nu/\nu_0)^{-q}$, $q\approx 0.88$, and $\nu_0=2.32\times 10^{4}$\,GHz. The diagonal line gives the best-fit theoretical relationship between the two quantities corresponding to $t_0=0.79$\,s, see text.
}\label{lags}
\end{figure}

We can solve for $b$, $p$, $\nu_0$, $F_0$, $\nu_{\rm max}$ and $t_0$ by only assuming $a=2 b$. From the measured fluxes, we obtain $b$, $p$, $\nu_0$, $F_0$ and $\nu_{\rm max}$ using Equations (\ref{integral}--\ref{F0}). We find the fitted spectrum is very insensitive to $\xi_{\rm min}(\nu)$, Equation (\ref{ximin}), as long as it is low enough. We can just use any low value of it, or just assume $\xi_{\rm min}=1$, and check a posteriori the consistency of the choice. Similar to \citet{Rodi21}, we find the presence of an additional hard component beyond $\nu_{\rm max}$, apparently due to the emission of an accretion disk. Given the limited range of the fitted frequencies, we fit it phenomenologically as a power law, $F_{\nu,\rm disk}=F_{\rm disk}(\nu/10^5{\rm GHz})^{\alpha_{\rm disk}}$. Then, using the obtained values of $b$, $p$ and $\nu_0$, we can fit the time lags using Equations (\ref{alpha_q}) and (\ref{t0}). However, with the MCMC technique, we fit the flux and time-lag measurements simultaneously. The fitted parameters and the correlations between them are shown in Figure \ref{basic_fits}, and the parameters are listed in Table \ref{basic}. The best-fitting values are given as the median of the resulting posterior distributions, and the lower and upper uncertainties are reported as the range between the median and the 15th percentile, and the 85th percentile and the median, respectively. These uncertainties correspond approximately to $1\sigma$ errors. We use these best-fit values as well as the best-fit values of $D$ and $i$ in our estimates in this subsection. 

Figure \ref{spec} shows the observed average radio-to-optical spectrum fitted by the above model. The best-fitting spectral indices in the optically thick and optically thin regimes are then $\alpha\approx 0.25$ and $\alpha_{\rm thin}\approx -0.61$, respectively. We show the theoretical spectrum calculated by integrating Equation (\ref{integral}) and the asymptotic optically thick and thin spectra of Equation (\ref{solution}) for this fit.

We then show the time lags in Figure \ref{lags}, where we plot the values of the measured $\Delta t_{\rm a}$ against the separation in the dimensionless units, $\xi$, using Equations (\ref{xi_nu}) and (\ref{t0}). At the best-fit values of $b$ and $p$, $q\approx 0.883$, see Equation (\ref{alpha_q}). The actual lags have to follow a single dependence relating the physical separation between the emission points to $\Delta t_{\rm a}$, which is shown by the diagonal line showing the linear relationship between $t_{\rm a}$ and $\Delta\xi_\nu$ corresponding to the best-fit value of $t_0$, see Equation \ref{t0}. We see a certain offset between the points corresponding to the lags between the sub-mm frequency of 343.5\,GHz and 6 radio frequencies (blue error bars), and the lags measured between the radio frequencies (red error bars). This may be related to the different methods used in \tet to determine those. On the other hand, the offset is significantly reduced for $q=0.8$, which value of $q$, however, is not compatible with $\alpha\approx 0.25$. This may indicate that the jet is more complex than we assume, e.g., either $\Gamma$, $\Theta$ or $b$ are not constant at $z\geq z_0$. 

Our formalism assumes the lags correspond to propagation of perturbations between different values of $z_\nu$ at the jet speed, $\beta$. With this assumption, we obtain $z_0$ as a function of $\Gamma$, see Equation (\ref{z0}), \begin{equation}
z_0 \approx (2.37\times 10^{10}\,{\rm cm})(t_0/0.79\,{\rm s})\beta\Gamma\delta.
\label{z0M}
\end{equation}

We then use the solutions obtained in Appendix \ref{general} assuming $\gamma_{\rm min}=3$, $k_{\rm i}=0$. However, we only know $\nu_{\rm max}$ rather than $\gamma_{\rm max}$, see Equation (\ref{nu_range}). Since the solutions depend on $\gamma_{\rm max}$ relatively weakly, we assume here the best-fit values of $B_0=10^4$\,G and $\Gamma=2.2$ obtained in Section \ref{mcmc} for $\gamma_{\rm min}=3$, which yield $\gamma_{\rm max}=125$, which we will use hereafter in this subsection. Using Equation (\ref{Theta_f}), we obtain at the best fit
\begin{equation}
\Theta\approx \frac{2.32\degr}{\left(\beta\Gamma\right)^{1.89}\delta^{2.57}\beta_{\rm eq}^{0.11}}.
\label{Theta}
\end{equation}
At $\beta\approx 1$ and $\beta_{\rm eq}=1$, $\Theta\approx 0.53\degr \Gamma^{0.67}$. Next, Equations (\ref{B_0ft}--\ref{n_0ft}) give at the best fit,
\begin{align}
&B_0\approx \frac{8.2\times 10^3\,{\rm G}(\beta\Gamma)^{0.22}} {\delta^{0.13}\beta_{\rm eq}^{0.22}},
\label{B0final}\\
&n_0\approx \frac{1.8\times 10^{12}\,{\rm cm}^{-3}(\beta\Gamma)^{0.43}\beta_{\rm eq}^{0.57}}{\delta^{0.26}},
\label{N0final}
\end{align}
with $B_0\propto \Gamma^{0.35}$ and $n_0\propto \Gamma^{0.70}$ at $\beta\approx 1$. Equation (\ref{Theta}) shows that we {\it cannot\/} determine both $\Theta$ and $\delta$ even assuming a value of $\beta_{\rm eq}$ (on which $\Theta$ depends very weakly). We can also calculate the Thomson scattering optical depth along the jet radius at $z\geq z_0$, which equals 
\begin{equation}
\tau_{\rm T}(\xi)=\sigma_{\rm T}n_0 f_N z_0 \tan\Theta \xi^{1-2 b}
\approx \frac{2.5\times 10^{-4}\beta_{\rm eq}^{0.46} }{(\beta\Gamma)^{0.46} \delta^{1.82}\xi^{1.2}}.
\label{tauT}
\end{equation}

At $i=64\degr$ and $\Gamma=2$, 3, 4, we have $\delta\approx 0.81$, 0.57, 0.43, and, at $\beta_{\rm eq}=1$, $\Theta\approx 1.4\degr$, $1.4\degr$, $1.5\degr$, $B_0\approx 1.0,\,1.1,\,1.2\times 10^4$\,G, $z_0\approx 3.3,\,3.8,\,4.0\times 10^{10}$\,cm, and $\tau_{\rm T}(\xi=1)\approx 2.9,\,4.3,\,6.1\times 10^{-4}$, respectively. The values of $z_0$ correspond to $\approx (2.8$--$3.3)\times 10^4 R_{\rm g}$ at an assumed $M=8\msun$. We find that $\Theta$, $B_0$ and $z_0$ depend relatively weakly on $\Gamma$ for $1.5\lesssim \Gamma\lesssim 5$. 

We determine the typical Lorentz factors, $\gamma_\nu$, of relativistic electrons giving rise to the emission at $\nu$, which in the partially self-absorbed regime originates mostly from $z_\nu$, see Equation (\ref{gamma_nu}). We obtain
\begin{equation}
\gamma_\nu \approx 32 \beta_{\rm eq}^{0.11} (\beta\Gamma)^{-0.11} \delta^{-0.43}(\nu/\nu_0)^{0.014}.
\label{av_gamma}
\end{equation}
In order to obtain a power-law emission in that regime, we need $\gamma_{\rm min}$ to be a factor of a few smaller. Thus, we require $\gamma_{\rm min}\lesssim 10$ for the validity of the model. The maximum $\gamma$ corresponds to the fitted $\nu_{\rm max}$, Equation (\ref{nu_range}). From that, we obtain $\gamma_{\rm max}$ ranging from $\approx$123 to 147 for $\Gamma$ increasing from 2 to 4. Combining this with the values of $\tau_{\rm T}$ from Equation (\ref{tauT}), we find that the power in the synchrotron self-Compton component is relatively similar to that in the synchrotron one, $P_{\rm SSC}\lesssim \tau_{\rm T}\gamma_{\rm max}^2 P_{\rm S}$.

\begin{figure}[t]
\centerline{
\includegraphics[width=7cm]{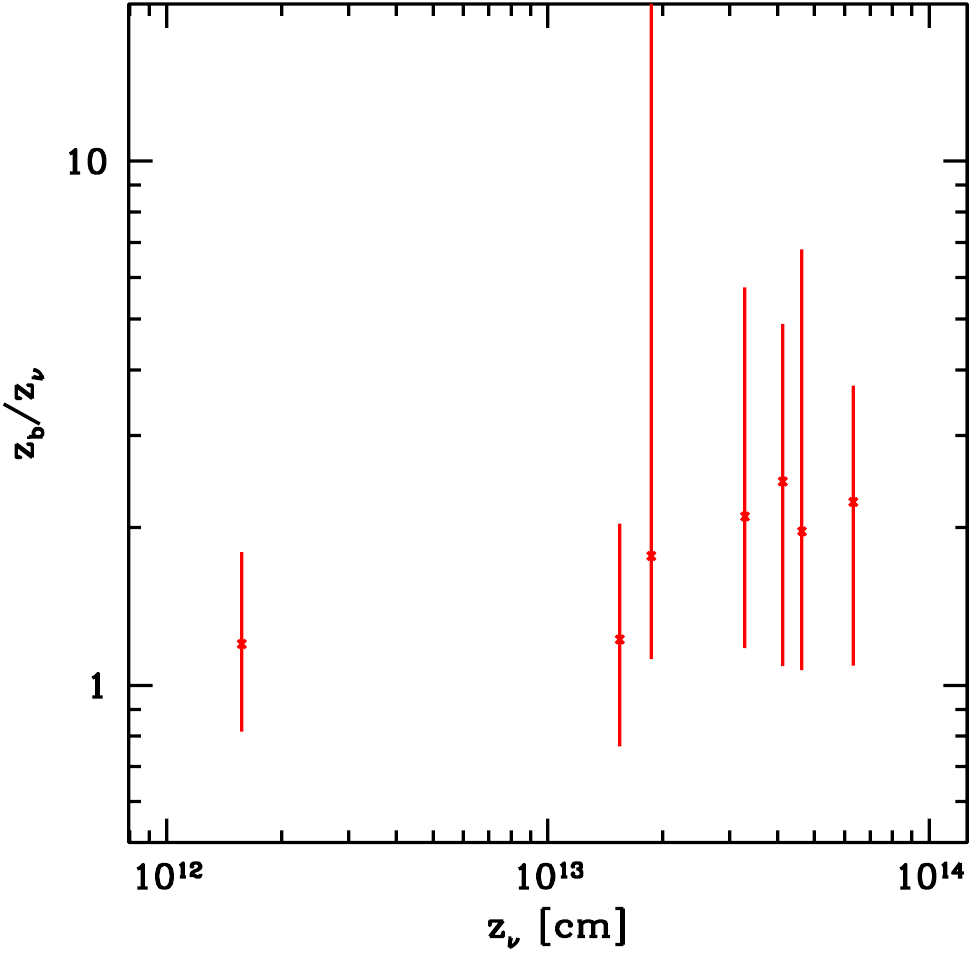}}
  \caption{The locations of the emission at the observed frequencies inferred from the break in the power spectra with the assumption of $z_{\rm b}=\beta c/f_{\rm break}$, shown as their ratio to the locations based on time lags and the slope of the partially self-absorbed spectrum, $z_\nu\approx z_0(\nu/\nu_0)^{-0.88}$  (with $z_0$ for $\Gamma= 3$ and $i=64\degr$).
}\label{z}
\end{figure}

We can then consider implications of the break frequencies, $f_{\rm b}$, in the power spectra for different frequencies measured by \tet. For those power spectra, most of the variability power per $\ln f$ occurs at $f\leq f_{\rm b}$, with the variability at higher frequencies strongly damped, see figs.\ 3 and 5 in \tet. We define the distance, $z_{\rm b}$, as that covered by a jet element moving with the jet velocity during the time\footnote{\tet assumed $z_\nu=z_{\rm b}\equiv\beta c\delta/f_{\rm b}(\nu)$, which they used as the final condition determining the jet parameters. Thus, they transformed the observed variability frequency to the {\it jet frame}, $f_{\rm b}/\delta$, and multiplied the resulting time scale, $\delta/f_{\rm b}$, by the jet velocity in the {\it observer's frame}, $\beta c$, which does not appear to be correct. We note that in the present case we consider the light curve originating from a {\it fixed\/} region of the jet around $z_\nu$. While the plasma in that region is moving, two adjacent maxima in the observed light curve are emitted from the same region in the frame connected to the BH, which is the same frame as the observer's one (in the absence of a redshift). Thus, a frequency inferred from the variability power spectrum should not be transformed.} $1/f_{\rm b}$, 
\begin{equation}
z_{\rm b}(\nu)\equiv\beta c/f_{\rm b} (\nu).
\label{zbreak}
\end{equation}
We can compare it to the distance along the jet from the BH up to the location of the peak emission at $\nu$, i.e., $z_\nu$ (Equations \ref{xi_nu}, \ref{z0}). In our model, $z_\nu\approx z_0(\nu/\nu_0)^{-0.88}$ with $z_0\propto \beta/(1-\beta\cos i)$, giving $z_{\rm b}/z_\nu\propto 1-\beta\cos i$. Then, this ratio depends only weakly on $\beta$ (or $\Gamma$); at $i=64\degr$, $1-\beta\cos i$ changes only from 1 at $\beta\ll 1$ to 0.56 at $\beta\approx 1$. This implies that this correlation cannot be used to determine the actual bulk Lorentz factor of the jet. 

Figure \ref{z} shows $z_{\rm b}/z_\nu$ vs.\ $z_\nu$ for $\Gamma= 3$. We see an approximately constant ratio of $z_{\rm b}/z_\nu\approx 1.5$--2. Therefore, $z_{\rm b}$ is proportional and close to the travel time along $z_\nu$ in all of the cases. A possible explanation of the damping of the variability at frequencies $>c/z_\nu$ appears to be a superposition of the contributions to the emission from different parts of the region dominating at a given $\nu$, which is $\propto z_{\nu}$, as shown in Figure \ref{spatial}. The peak of ${\rm d}F_\nu/{\rm d}\,\ln z$ for $p=2.21$ is at $\approx 1.15 z_\nu$ and its width defined by ${\rm d}F_\nu/{\rm d}\,\ln z$ decreasing to the 50\% of the peak is $(0.65$--$3.16) z_\nu$. Thus, if different parts vary independently, the variability will be damped at $f\gtrsim c/(2 z_\nu)$, as observed.

Alternatively, the observed radio/IR variability can be driven by the variable power supplied from the vicinity of the BH with a wide range of frequencies \citep{Malzac13,Malzac14} and then transferred upstream. Then, the travel time can act as a low-pass filter, removing most of the variability at frequencies $f>\beta c/z_\nu$. This can happen due to damping of perturbations along the jet due to some kind of internal viscosity, e.g., collisions between shells within the jet moving with a range of velocities \citep{Jamil10}. The process would be then analogous to viscous damping in accretion disks, where modulations with a period shorter than the signal travel time across the disk are strongly damped \citep{ZKM09}. This picture is also compatible with the integrated fractional variability of the power spectra (RMS) decreasing with the decreasing $\nu$ (as shown in fig.\ 5 of \tet). This means increasing the distance travelled along the jet leads to the increasing damping. 

We note that the break frequencies in the power spectra of \tet have been defined by choosing a specific, and not unique, algorithm, as well as the obtained values of $f_{\rm b}$ are close to the minimum frequency at which the power spectrum is measured for $f<10$\,GHz, which limits the accuracy of the determination of those $f_{\rm b}$. Also, while the damping of variability above $\beta c/z_\nu$ clearly occurs, details of the physics behind it remain uncertain, and the damping could start at $f\sim \beta c/(2 z_\nu)$ instead of exactly $\beta c/z_\nu$. Summarizing, our results are completely compatible with the variability damping at time scales shorter than the light/jet travel time across $z_\nu$. However, unlike our previous estimates from the observed spectrum and time lags, which are based on a relatively rigorous and well-understood model, the detailed cause of the connection between the break frequencies and the distance along the jet remains uncertain.

We can also consider the prediction of the location of the bulk of the 15\,GHz emission, $z_\nu\approx z_0(\nu/\nu_0)^{-0.88}\approx 2.5\times 10^{13}$\,cm (at $\Gamma=3$, but weakly dependent on it), with the jet angular size at this frequency from the VLBA observation on 2018 March 16 (MJD 58193), reported in \tet as $0.52\pm0.02$\,mas. the deprojected size is $(2.60\pm 0.10)\times 10^{13}$\,cm. The total flux density at 15 GHz was measured as $F_\nu \approx 20.0\pm 0.1$\,mJy. However, the VLBA observation was 27\,d before the radio/sub-mm ones. On MJD 58220, our best-fit spectral model yields $F_\nu \approx 56\pm 1$\,mJy. Within the framework of the continuous conical jet model, we have $z_\nu\propto F_\nu^{(p+6)/(2p+13)}$ (Equation \ref{z_0}; \citealt{ZLS12}). Thus, for $p=2.2$ we predict the size at 15 GHz on MJD 58220 being $(56/20)^{0.47}\approx 1.6$ times larger than that on MJD 58193, namely $\sim 4\times 10^{13}$\,cm. While somewhat larger than the above $z_\nu$, this size appears consistent with it since the peak of ${\rm d}F_\nu/{\rm d}\,\ln z$ for $p=2.21$ is at $\approx 1.15 z_\nu \approx 3.0\times 10^{13}$\,cm, and that spatial distribution is broad and skewed toward higher distances, see Figure \ref{spatial} and the discussion of it above.

We then estimate the rate of pair production. For \source, pair production within the hot plasma was calculated by \citet{Zdziarski21c} based on the spectrum observed by \integral in the hard state. That spectrum was measured up to $\sim$2\,MeV, well above the pair production threshold of 511\,keV, and modelled by Comptonization. It was found that an appreciable pair abundance can be obtained only provided the hard X-ray source size is as small as several $R_{\rm g}$, while the spectroscopy based on the relativistic broadening of the fluorescent Fe K$\alpha$ line indicates a size of $\gtrsim 20 R_{\rm g}$. Then, the pair abundance within the Comptonizing plasma is very low. 

However, as discussed in Section \ref{pairs}, pair production within the jet base can be much more efficient. To calculate it, we adapt the results of \citet{Zdziarski21c}. We modify their equation (1) to calculate the photon density above the hot disk, dividing the total rate of the photon emission by $2\pi R_{\rm hot}^2$ (including both sides). We then use this photon density in equation (3) of that paper for the spectral parameters of average spectrum (table 2 in \citealt{Zdziarski21c}). This gives the pair production rate per unit volume. With the assumptions as in Section \ref{pairs}, we have 
\begin{equation}
2\dot N_+\approx 4.65\times 10^{40}\,{\rm s}^{-1}\left(\frac{R_{\rm hot}}{20 R_{\rm g}}\right)^{-3}\left(\frac{R_{\rm jet}}{10 R_{\rm g}}\right)^2.
\label{Nplus}
\end{equation}
This is then balanced by the sum of the rates of pair annihilation and pair advection. Using formulae in \citet{Zdziarski21c}, we have found that pair annihilation can be neglected for the advection velocity of $\beta_\pm\gtrsim 0.1$. It appears that such a velocity can be achieved due to the net momentum component of the pair-producing photons along the $z$ axis, see Figure \ref{jet_pairs}, and due to pair acceleration by radiation pressure of the disk photons \citep{B99_pairs}. Thus, while some of the produced pairs will annihilate (and a small fraction will be advected to the BH), a major fraction of the produced pairs will have a sufficient net bulk velocity to escape upstream.

Then, the lepton flow rate through the jet, Equation (\ref{ndot_e}), for $\gamma_{\rm min}=3$ is 
\begin{equation}
\dot N_{\rm e}\approx \frac{6.7\times 10^{40}{\rm s}^{-1}\beta_{\rm eq}^{0.35}}{(\beta\Gamma)^{0.35}\delta^{3.39}}
\propto\Gamma^{3.05},
\label{Nrel}
\end{equation}
where Equations (\ref{z0M}), (\ref{Theta}), (\ref{N0final}) have been used and the proportionality assumes $\beta\approx 1$. Comparing with Equation (\ref{Nplus}), we find $\dot N_{\rm e}>2 \dot N_+$ at any $\Gamma$ for $R_{\rm hot}=20 R_{\rm g}$, $R_{\rm jet}=10 R_{\rm g}$ and $\gamma_{\rm min}=3$. Thus, at these parameters the synchrotron-emitting plasma is never composed of pure pairs. If we assume either $R_{\rm jet}=15 R_{\rm g}$ or $\gamma_{\rm min}=10$, we find $\dot N_{\rm e}=2 \dot N_+$ at $\Gamma\approx 2$, which thus represent the minimum possible $\Gamma$ for these parameters. While the hot disk and jet radii and $\gamma_{\rm min}$ are poorly constrained, we consider the fact that the numbers in Equations (\ref{Nplus}) and (\ref{Nrel}), obtained with completely different physical considerations, are of the same order of magnitude, to be highly remarkable and indicating that indeed the two rates may be similar in this source. Then, the jet can contain a large fractional abundance of pairs, and they can dominate by number over the ions.

The pairs produced in the jet base and then advected to large distances will eventually leave the jet and enter the ISM. Our estimated rate, $\sim\! 10^{40}$--$10^{41}$\,s$^{-1}$, is lower than the approximate estimates of the positron production rates in microquasars of \citet{Guessoum06}. It is also a few orders of magnitude below the total rate of pair annihilation in the Galaxy, which has been estimated by \citet{Siegert16} as $\approx\! (3$--$6)\times 10^{43}$\,s$^{-1}$. With the past and present X-ray monitors (ASM, \citealt{ASM}; MAXI, \citealt{Matsuoka09}; BAT, \citealt{BAT}), we can detect all of the outbursting accreting BH binaries with luminosities of more than a few percent of the Eddington luminosity. Based on the MAXI data, there is $\sim$0.5 source in outburst at given time (the MAXI team, private communication). Thus, the average contribution of such sources to the Galactic positrons appears to be negligible, contrary to earlier estimates \citep{Guessoum06, Weidenspointner08}.

\begin{figure*}
\centerline{
\includegraphics[width=1.05\textwidth]{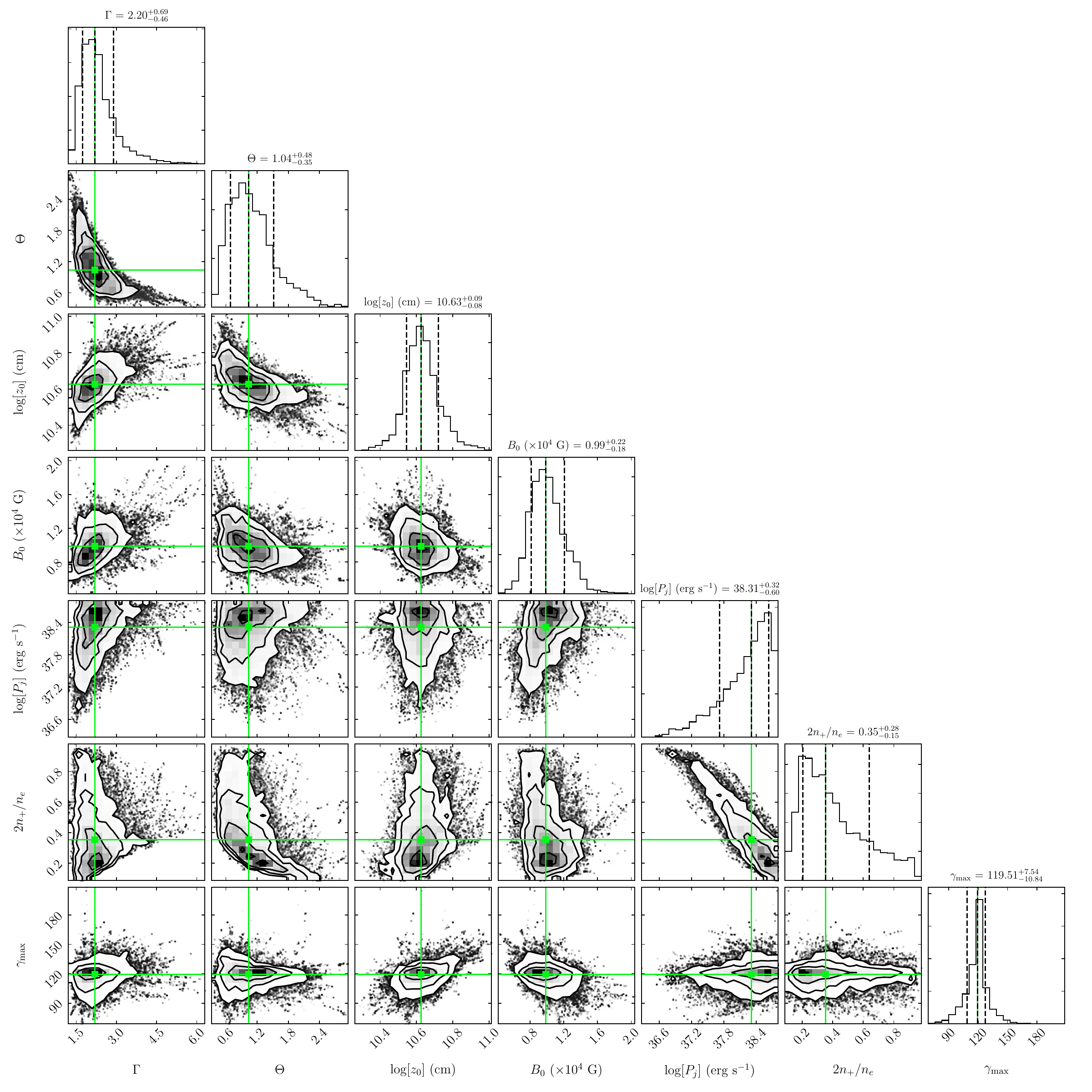}}
  \caption{(a) The MCMC fit results for $\Gamma$, $\Theta$, $z_0$, $B_0$, $P_{\rm j}$, $2n_+/n_{\rm e}$ and $\gamma_{\rm max}$ assuming $\gamma_{\rm min}=3$ and $\epsilon_{\rm eff}=0.3$. The meaning of the panels and lines is the same as in Figure \ref{basic_fits}. See Section \ref{mcmc} for details. 
}\label{fits_full}
\end{figure*}

\setcounter{figure}{7}
\begin{figure*}
\centerline{
\includegraphics[width=1.05\textwidth]{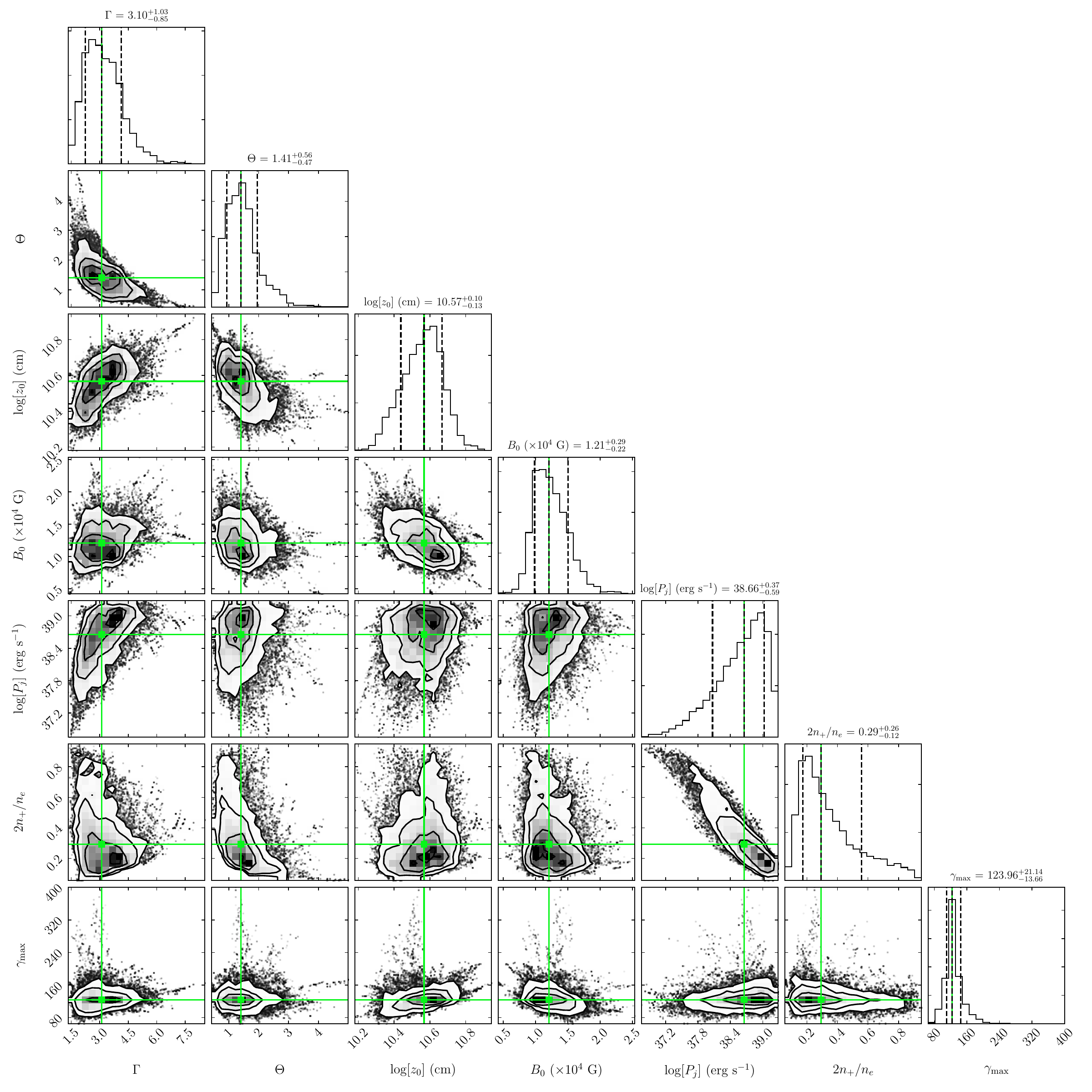}}
  \caption{(b) The MCMC fit results for $\gamma_{\rm min}=10$ and $\epsilon_{\rm eff}=0.1$.
}
\end{figure*}

\begin{table*}\centering
\caption{The parameters of the jet in \source other than those given in Table \ref{basic}. }
\label{rest}
\begin{tabular}{cccccccc}
\hline
$\gamma_{\rm min}$ & $\epsilon_{\rm eff}$ & $\Gamma$ & $\Theta$ & $\log_{10} z_0$ & $B_0$ & $\log_{10} P_{\rm j}$ & $\gamma_{\rm max}$\\
 &  & & $\degr$ & cm & $10^4$\,G & erg\,s$^{-1}$ & \\
\hline
3f & 0.3f & $2.20^{+0.69}_{-0.46}$ & $1.04_{-0.35}^{+0.48}$ & $10.63^{+0.09}_{-0.08}$ & $0.99^{+0.22}_{-0.18}$ & $38.31^{+0.32}_{-0.60}$ & $120^{+8}_{-11}$ \\
10f &0.1f & $3.10^{+1.03}_{-0.85}$ & $1.41_{-0.56}^{+0.47}$ & $10.57^{+0.10}_{-0.13}$ & $1.21^{+0.29}_{-0.22}$ & $38.66^{+0.37}_{-0.59}$ & $124^{+21}_{-14}$ \\
\hline
\end{tabular}
\tablecomments{The fits are done with the MCMC method for the assumed the fixed (marked by 'f') minimum electron Lorentz factor, $\gamma_{\rm min}$, and the accretion efficiency, $\epsilon_{\rm eff}$.}
\end{table*}

Next, we calculate the jet power. The power in the relativistic electrons and magnetic fields, Equation (\ref{Prel}), becomes at $z_0$
\begin{equation}
P_B+P_{\rm e}\approx 1.9\times 10^{36}{\rm erg\,s}^{-1} \frac{3+2\beta_{\rm eq}}{6\beta_{\rm eq}^{0.65}} \frac{\Gamma^{0.65}}{
 \beta^{0.35}\delta^{3.39}}.
\label{Prel2}
\end{equation}
which increases very fast with $\Gamma$, approximately as $\propto \Gamma^4$ at $\beta\approx 1$. At $\beta_{\rm eq}=1$, $\Gamma=3$, this power is $\approx 2.2\times 10^{37}$\,erg\,s$^{-1}$. The power associated with the bulk motion of cold matter, Equations (\ref{Pcold}), (\ref{Pcold3}), is 
\begin{align}
&P_{\rm i}\approx 1.2\times 10^{38}{\rm erg\,s}^{-1}(\Gamma-1)\times\\
&\left[
\frac{\beta_{\rm eq}^{0.35}}{ (\beta\Gamma)^{0.35}\delta^{3.39}}-0.7\left(\frac{R_{\rm hot}}{20 R_{\rm g}}\right)^{-3}\!\!\left(\frac{R_{\rm jet}}{10 R_{\rm g}}\right)^2\right]\!.\nonumber
\end{align}
The first term is approximately $\propto \Gamma^3(\Gamma-1)$. 

To constrain $P_{\rm j}$ by the accretion power, we use the estimate of the hard-state bolometric flux of $F_{\rm bol}\approx 1.4\times 10^{-7}$\,erg\,cm$^{-2}$\,s$^{-1}$ \citep{Shidatsu19}. This yields $L\approx 1.5(D/2.96\,{\rm kpc})^2 10^{38}$\,erg\,s$^{-1}$ and
\begin{equation}
\dot M c^2 \approx 1.5\times 10^{39}\left(D\over 2.96\,{\rm kpc}\right)^2 \left(\epsilon_{\rm eff}\over 0.1\right)^{-1} \,{\rm erg\,s}^{-1}. 
\label{Mdotc2}
\end{equation}
For the default parameter values, $P_{\rm j}\lesssim \dot M c^2$ implies $\Gamma\lesssim 3.3$. If pair production is efficient enough, we also have a lower limit on $\Gamma$ from the requirement of $P_{\rm i}> 0$. The allowed range depends significantly on the assumed parameters, in particular $\gamma_{\rm min}$, $R_{\rm hot}$ and $R_{\rm jet}$. E.g., at $\gamma_{\rm min}=10$, $R_{\rm hot}=20 R_{\rm g}$ and $R_{\rm jet}=10 R_{\rm g}$, $\Gamma\gtrsim 2.4$ is required.

We can then compare the total jet power, $P_{\rm j}$, with the synchrotron power. At the low $\gamma_{\rm max}$ implied by the $\nu_{\rm max}$ fitted to the spectrum, we find $P_{\rm S}\ll P_{\rm j}$ always. For example, $P_{\rm S}\approx 0.009 P_{\rm j}$ at the maximum allowed $\Gamma\approx 3.3$, and $P_{\rm S}\approx 0.02 P_{\rm j}$ at $\Gamma=2$. On the other hand, we have found $P_{\rm S}\sim 0.5 (P_B+P_{\rm e})(z_0)$, weakly depending on either $\Gamma$ or $\gamma_{\rm min}$. Thus, the synchrotron emission can be entirely accounted for by the power in electrons and magnetic fields at $z_0$, and most of the decline of $P_B+P_{\rm e}$ with the distance can be due to the synchrotron losses. However, we may see that the decline of $(P_B+P{\rm e})$ with $\xi$ is slower than that of the synchrotron power. If the former would be just to the synchrotron emission, we would have ${\rm d}(P_B+P{\rm e})/{\rm d}\xi + {\rm d}P_{\rm S}/{\rm d}\xi =0$, while the former and the latter terms are $\propto -\xi^{1-2 b}$ and $\propto \xi^{2-4 b}$. This implies either some electron re-acceleration at $z>z_0$ at the expense of $P_{\rm i}$, or more complexity of the actual physical situation, with the initial energy loss in the flow being faster and followed by a slower one. 

In the framework of models with the jet dissipation mechanism being the differential collimation of poloidal magnetic surfaces, the obtained $\Theta\Gamma\ll 1$ indicate the jet magnetization at $z\gtrsim z_0$ is low. Using Equation (\ref{opening}), we have (at $\beta\approx 1$)
\begin{equation}
\sigma=(\Theta\Gamma/s)^2 \approx \frac{8.4\times 10^{-5} \Gamma^{3.35}}{\beta_{\rm eq}^{0.22} s^{2}}.
\label{sigma_n}
\end{equation}
At $\beta_{\rm eq}=1$ and assuming $s=0.6$ (as found as the average value for a large sample of radio-loud AGNs by \citealt{Pjanka17}), we obtain $\sigma\approx 0.0093(\Gamma/3)^{3.35}$. This can be compared to $\sigma$ from its definition, Equation (\ref{sigma}), which equals,
\begin{equation}
\sigma\approx \beta_{\rm eq}^{-1}\left[2/3+130 (1 -2 n_+/n_{\rm e})\right]^{-1}.
\label{sigma_Brho}
\end{equation}
In the absence of pairs, $\sigma\approx 0.0078/\beta_{\rm eq}$. Comparing the two estimates of $\sigma$, we see it requires $\Gamma\gtrsim 3$ at $\beta_{\rm eq}=1$. However, the actual value of $s$ is uncertain, there could be ions associated with background electrons piled up at $\gamma<\gamma_{\rm min}$, and, importantly, $\beta_{\rm eq}$ could be $\gg 1$. Still, the low magnetization implied by Equation (\ref{sigma_n}) disfavors the case of strong pair dominance, $(1-2 n_+/n_{\rm e})\ll 1$.

Using $\sigma\ll 1$, we can calculate the magnetic fluxes in the model with extraction of the BH rotational power. The jet magnetic flux from Equation (\ref{phi_jet}) with $z_0 B_0$ from Equations (\ref{z0M}) and (\ref{B0final}) is then 
\begin{equation}
\Phi_{\rm j}\approx (4.1\times 10^{21}\,{\rm G\,cm}^{2}) \frac{s[1+(1-a_*^2)^{\frac{1}{2}}] (\beta\Gamma)^{1.22}\delta^{0.87}}{(\ell/0.5)a_*\beta_{\rm eq}^{0.22}},
\label{Phi_jet2}
\end{equation} 
which is $\approx 5.3\times 10^{21}\,{\rm G\,cm}^{2}$ for $a_*=1$, $\Gamma=3$, $\ell=0.5$, $s=0.6$, $\beta_{\rm eq}=1$. The flux threading the BH, Equation (\ref{phi_BH}) with $\dot M$ estimated as above from $L$, is
\begin{equation}
\Phi_{\rm BH}\approx (1.3\times 10^{22}\,{\rm G\,cm}^{2}) \frac{\phi_{\rm BH}}{50}\frac{D}{3\,{\rm kpc}} \left(\frac{\epsilon_{\rm eff}}{0.1}\right)^{-1/2},
\label{Phi_BH2}
\end{equation}
where $M=8\msun$ was assumed for both ($\Phi\propto M$). At $\phi_{\rm BH}=50$ and the assumed parameters, the two fluxes are approximately equal for $a_*\approx 0.7$. We consider the close agreement of the above two estimates to be very remarkable. They are based on completely different physical considerations. Thus, our results are consistent with the jet being powered by the BH rotation and the accretion flow being magnetically arrested. In this case, the jet power is maximal and given by Equation (\ref{P_tot}). However, we have found that if pairs dominate in the jet, $P_{\rm j}\ll \dot M c^2$. This requires either $a_*\ll 1$, the magnetic field in the flow is weaker than that in a MAD, or that some assumption in the model with extraction of the BH rotation power, e.g., the ideal MHD, are not satisfied.

\subsection{Numerical estimates}
\label{mcmc}

In order to solve directly for the physical jet parameters and their uncertainties, we use again the MCMC method. In the fits shown in Figure \ref{basic_fits}, we fitted $b$, $p$, $\nu_0$, $\nu_{\rm max}$, $F_0$, $t_0$, $F_{\rm disk}$ and $\alpha_{\rm disk}$ with the minimum assumption of $a=2 b$, and, in particular, without the need to specify the value of $\Gamma$. Now we fit for all of the parameters. However, since the solution given in Appendix \ref{general} is given in terms of $\gamma_{\rm max}$ rather than $\nu_{\rm max}$, we fit for the former (which yields $\nu_{\rm max}$ given the values of $\Gamma$, $i$ and $B_0$, see Equation \ref{nu_range}). 

In particular, we determine $\Theta$ from Equation (\ref{Theta_f}), $z_0$ from Equation (\ref{z0}) and $B_0$ from Equation (\ref{B_0ft}). That requires specifying $\Gamma$ (which is then a free parameter) and $\gamma_{\rm min}$. We fix $\beta_{\rm eq}=1$ and $k_{\rm i}=0$. However, in order to be able to constrain $\Gamma$ rather than have it entirely free, we include further constraints, using the pair production rate of Equation (\ref{Nplus}) and requiring $2 \dot N_+/\dot N_{\rm e}\leq 1$ in Equation (\ref{Pcold3}) and from the maximum possible jet power, $P_{\rm j}\leq \dot M c^2$, Equations (\ref{Prel}--\ref{Pjet_dotM}). These constraints require specifying $R_{\rm hot}$ and $R_{\rm jet}$, the bolometric luminosity, $L$, and the accretion efficiency, $\epsilon_{\rm eff}$. We then solve simultaneously for all of the parameters, including $b$, $p$, $\nu_0$, $F_0$, $t_0$, $F_{\rm disk}$ and $\alpha_{\rm disk}$. Those parameters have now values similar to those shown in Figure \ref{basic_fits}, and we thus do not show them again. 

In the solution, we sample $D$ and $i$ as described at the beginning of Section \ref{MAXI}. We assume $L=1.5\times 10^{38}$\,erg\,s$^{-1}$, $X=0.7$, $R_{\rm hot}=20 R_{\rm g}$, $R_{\rm jet}=10 R_{\rm g}$ (for $M=8\msun$). We show the resulting posterior distributions for two cases with ($\gamma_{\rm min}=3$, $\epsilon_{\rm eff}=0.3$), and with ($\gamma_{\rm min}=10$, $\epsilon_{\rm eff}=0.1$), in Figures \ref{fits_full}(a), (b), respectively, and list the fitted parameters in Table \ref{rest}. We see that the obtained ranges of $\Gamma$ and $\Theta$ depend on those two sets of assumptions, being larger for for the latter case. The allowed maximum jet power is $\propto \epsilon_{\rm eff}^{-1}$, and then it is higher in case (b). On the other hand, the obtained values of $z_0\approx 2$--$4\times 10^{10}$\,cm and $B_0\approx 10^4$\,G depend relatively weakly on those assumptions. For the sake of brevity, we have not shown the effect of changing the values of $R_{\rm hot}$ and $R_{\rm jet}$. For example, for $R_{\rm hot}>20 R_{\rm g}$, pair production will be less efficient, which would in turn allow fewer leptons in the flow and lower values of $\Gamma$, see Equations (\ref{Nplus}--\ref{Nrel}). Thus, we cannot conclusively rule out values of $\Gamma\lesssim 1.5$. Then, values of $\Gamma$ higher than those obtained above would be possible for $\epsilon_{\rm eff}<0.1$. 

Figures \ref{fits_full}(a--b) also show $\gamma_{\rm max}$ and the pair abundance, $2n_+/n_{\rm e}$. The former ir relatively tightly constrained in the $\approx 110$--150 range. The latter is strongly anticorrelated with the jet power, being low at the maximum $P_{\rm j}$ and close to unity at the minimum $P{\rm j}$, in agreement with our considerations in Section \ref{analytical}. We find the synchrotron power, Equation (\ref{P_S}), is typically $P_{\rm S}\sim 0.01 P_{\rm j}$, as in Section \ref{analytical}, and thus the jet radiative efficiency, $P_{\rm S}/P_{\rm j}$, is low. In our fits, we have not used constraints from the break frequencies in the power spectra and from the jet spatial extent measurement, following our discussion in Section \ref{analytical}.

\section{Discussion}
\label{discussion}

\subsection{The location of the dissipation zone}
\label{dissipation}

Our results indicate that the jet synchrotron emission, and thus electron acceleration, starts at the distance of $z_0\sim\! 3\times 10^4 R_{\rm g}$ away from the BH. This is similar to the situation in blazars, in which the so-called blazar zones are found to be at distances between that of the broad-line regions and of the molecular torii, i.e., $z_0\sim\! 10^3$--$10^5 R_{\rm g}$ (\citealt{MS16} and references therein). Then, a radio-optical lag measured in the blazar BL Lac yields $z_0$ of several times $10^4 R_{\rm g}$ \citep{Marscher08}. The region upstream to the onset of the emission is called an acceleration and collimation zone (ACZ). This similarity of $z_0/R_{\rm g}$ in jets in blazars and in binaries accreting onto BHs is consistent with the scale invariance of relativistic jets \citep{HS03}. The absence/weakness of emission closer to the BH is also consistent with the jets formed by magnetic processes and being initially Poynting-flux dominated \citep{BZ77, BP82, Blandford19}. The onset of the electron acceleration is often associated with the presence of a standing shock. For example, \citet{Ceccobello18} invoke recollimation shocks at a fast magnetosonic point. A requirement for a formation of a shock is $\sigma<1$, which agrees with our estimates of $\sigma\ll 1$ in the synchrotron emission region, see Equations (\ref{sigma_n}--\ref{sigma_Brho}). An alternative explanation for the presence of an ACZ in BH X-ray binaries invokes the colliding shell model, in which the dissipation region is associated with the distance at which the shells begin to collide \citep{Malzac13,Malzac14}.

Our determination of $z_0$ is within that found for the jet of the BH X-ray binary MAXI J1836--194 of $2\times 10^3$--$10^6 R_{\rm g}$ in the model of \citet{Lucchini21}. On the other hand, our values of $z_0$ are significantly larger than that of $z_0\sim 10^3 R_{\rm g}$ inferred from lags of the IR/optical emission with respect to X-rays measured in the BH X-ray binaries GX 339--4 and V404 Cyg \citep{Gandhi08, Gandhi17, Casella10}. We point out that the analysis of \tet found the cross-correlation between the optical light curve ($3.9\times 10^5$\,GHz) and that of X-rays (shown in their fig.\ 8) to be rather complex in \source. The dominant feature was an anticorrelation centered on the zero lag, on top of which there was a much weaker positive correlation with the peak at the optical vs.\ X-rays of $150^{+500}_{-700}$\,ms, and a rather low relative amplitude of the correlation peak of $\approx$0.03, in contrast wit the IR/optical rms variability of a few tens of per cent (\tet). Given that weakness and the very large error bar on the lag, we have not used that constraint in our modelling. The optical/X-ray cross correlation in \source was later studied in more detail by \citet{Paice21}. They found that the cross-correlation averaged over 2-s segments of the light curves shows the positive correlation to dominate, with the optical lag of $\approx$150--200\,ms. Curiously, this lag was almost the same for all six epochs they studied, in spite of a spread of the X-ray flux up to a factor of three (see fig.\ 1 in \citealt{Paice21}).

An interpretation of the positive lag of the optical emission with respect to X-rays as due to the signal propation from the BH vicinity to the region of the onset of the jet dissipation implies $t_0\approx 150$--200\,ms. Thus, this region would lie at a significantly lower distance than that found in our study, where $t_0\approx 790_{-190}^{+300}$\,ms, see Figure \ref{basic_fits}. We have found that including this constraint would worsen our fit to the remaining observables, especially to the 343.5\,GHz flux. Given the weakness of the correlation, the complexity of the cross-correlation shape and the unexplained constancy of the positive lag component, we consider that alternative interpretation to be uncertain. A model including that lag would have to also account for the dominant anticorrelation, possibly due to the effect of synchrotron cooling on the X-ray spectrum emitted by the accretion flow \citep{VPV13}, which is beyond the scope of the present paper.

\subsection{Electron energy losses and re-acceleration}
\label{losses}

We have parametrized the electron distribution as a power-law function of the distance, and assume that distribution keeps a constant shape. Such a situation requires the electron energy losses are moderate and satisfying $\dot\gamma \propto \gamma$. We compare here the time scale for synchrotron energy losses,
\begin{equation}
t_{\rm syn}=\frac{6\pi m_{\rm e}c\xi^{2 b}}{\sigma_{\rm T}B_0^2\gamma},
\label{t_syn}
\end{equation}
with the adiabatic/advection time scale,
\begin{equation}
t_{\rm ad}=\frac{3 z_0\xi}{2\beta\Gamma c}
\label{t_ad}
\end{equation}
(e.g., \zdz). We consider the solution in Section \ref{analytical} for $\Gamma=3$. At $\gamma\approx 30$, which corresponds to the bulk of the partially self-absorbed emission, $t_{\rm syn}$ is shorter than $t_{\rm ad}$ for $\xi\lesssim 3$, and it is $\approx$3 times shorter at $z_0$. This implies that electrons responsible for the optically-thin part of the synchrotron emission have to be re-accelerated above $z_0$.

Calculating the electron distribution self-consistently as a function of the distance as well as accounting for the slope of the spectrum at $\nu<\nu_0$ is relatively complex, involving solving a kinetic equation with both losses and spatial advection (e.g., \zdz). This also requires taking into account losses from Compton scattering of synchrotron photons as well as the reduction of the electron energy loss rate due to self-absorption \citep{GGS88,KGS06}. Such a model is beyond the scope of the present work.

\subsection{Comparison with other jet models of accreting black holes}
\label{comparison}

The main independent study of the hard-state jet of \source is that by \citet{Rodi21}. They had at their disposal only the spectral data. They assumed $R/z=0.1$, corresponding to $\Theta=5.7\degr$, which is much larger than that found by us. They assumed $\Gamma=2.2$ following the result of \citet{Bright20} for the ejection during the hard-to-soft transition, but we note that $\Gamma$ of the hard-state jet is likely to be different. The jet model of \citet{Rodi21} is also different from ours, and considers an initial acceleration event followed by synchrotron cooling assuming no adiabatic losses (following \citealt{Peer09}). They do not show the spatial structure of their jet model, and thus we are not able to check whether that model would agree with our time-lag data. Still, they obtain relatively similar values of the distance of the onset of electron acceleration, $z_0\approx 2.8\times 10^{10}$\,cm, and the magnetic field strength at that distance, $B_0\approx 1.8\times 10^4$\,G. 

The very long time lags found in \tet unambiguously show that the radio/sub-mm emission originates at size scales several orders of magnitude higher than $R_{\rm g}$. The time lags between $\nu_1$ and $\nu_2$ are found to be approximately proportional to $\nu_2^{-1}-\nu_1^{-1}$. Knowing the break frequency, $\nu_0$, above which the entire synchrotron emission is optically thin, we can extrapolate this correlation and find the location corresponding to $\nu_0$. This is found to be $z_0\sim 3\times 10^4 R_{\rm g}$, with the uncertainty of a factor of at most a few. This rules out jet models predicting the onset of the synchrotron emission to in an immediate vicinity of the BH, for example that described in \citet{Giannios05} (based on the model of \citealt{Reig03}).

The model of \citet{Reig03} was developed in order to explain time lags of harder X-rays with respect to softer ones by Compton scattering. For that reason, the authors invoke a rather massive and extended jet, where multiple scattering of disk photons place. Consequently, this model (further developed in a number of subsequent papers of those authors) requires a rather large rate of the electron flow. For the parameters of \citet{Giannios05} (similar to those in \citealt{Reig03}), the base of the jet has a radius of $R_0=100 R_{\rm g}$ with the electron density of $n_{\rm e,0}\approx 3.43 \times 10^{16}$\,cm$^{-3}$ (for $M=8\msun$), which electrons flow upward with $\beta=0.8$. For matter with cosmic composition, this corresponds to the mass flow in the both jets of $2.4\times 10^{20}$\,g/s. On the other hand, the bolometric $L$ estimated for \source corresponds to $\dot M\approx 1.7(D/2.96\,{\rm kpc})^2(\epsilon_{\rm eff}/0.1)^{-1} 10^{18}$\,g/s, i.e., two orders of magnitude less, which rules out this case. We can also consider the case in which the leptons in the flow are pairs. In that case, $2\dot N_+ = 2 \pi R_0^2 n_{\rm e,0}^2 \beta \Gamma c \approx 1.2 \times 10^{44}$\,s$^{-1}$. This is 3--4 orders of magnitude higher than the pair production rate in this source calculated in Equation (\ref{Nplus}), which rules out this case too. Since the hard state of \source is rather similar to that of other X-ray binaries with accreting BHs, we find we can rule out this model in general. 

\subsection{Other constraints and caveats}
\label{caveats}

Our model is based on that of \citet{BK79} and \citet{Konigl81}, and it assumes uniform scaling of the emission regions, through the coefficients $a$ and $b$. As we see in Figure \ref{spec}, this model does not account for the observed flux at 339\,MHz, measured by \citet{Polisensky18}. This hints for the decline of the energy content in the relativistic electrons and magnetic field being initially faster (responsible for the emission closer to $z_0$) and then slower (responsible for the emission farther away from $z_0$). This would introduce more complexity in the modelling, and is beyond the scope of this work. On the other hand, the flux at 339\,MHz could be due to another component, in particular a pair of radio lobes at the jet ends. An assumption of our model is that the bulk of the emission at a given distance in the partially self-absorbed part of the spectrum occurs at a $\nu$ corresponding to $\tau\approx 1$. As we have found out, this corresponds to the synchrotron emission by electrons with $\gamma\sim 30$. If the minimum Lorentz factor of the electron distribution were higher, $\gamma_{\rm min}>30$, then the emission at a given distance in that part of the spectrum would be dominated by the electrons at $\gamma_{\rm min}$ instead, with no contribution from self-absorption.

We assumed the jet is already fully accelerated at $z_0$ and then does not decelerate. This may be not the case, and the available data do not exclude that. The jet model of \zdz allows for a variable $\Gamma$, and we could use some parametrization of $\Gamma(z)$ and refit our data (as done in \citealt{Zdziarski19c} for another source). This would, however, introduce more free parameters, and make the resulting fits less constrained than in the present case. We have also considered the steady state, while variability has been observed. However, the fractional variability was $\sim 0.3$ at the sub-mm range and much less than that in the radio regime. Thus, the variability can be considered as a small perturbation of the steady state. 

We also use a $\delta$-function approximation to the synchrotron process, which is a good approximation for power-law parts of the spectra, but becomes less accurate at cutoffs, given the single-electron synchrotron spectrum is quite broad \citep{GS65}. We assume the synchrotron emission of a single electron is isotropic in the plasma frame, which is valid for a tangled magnetic field, while we assume a toroidal field in some of our equations. Furthermore, we assume a sharp cutoff in the electron distribution at $\gamma_{\rm max}$. While this is not realistic, the actual form of the cutoff depends on details of the acceleration process and is poorly constrained. Thus, our determination of $\gamma_{\rm max}$ based on the observed cutoff in the optical range is only approximate.

Then, we have used our self-consistent set of equations, in which the slope of the partially self-absorbed part of the synchrotron spectrum is connected to the rate of decline of the energy density along the jet. The latter determines the relationship between the characteristic emitted frequency and the distance (Equation \ref{xi_nu}), and thus the time-lag vs.\ frequency relation. A significant discrepancy between the spectral slope and time lags vs.\ frequency was found in Cyg X-1 \citep{Tetarenko19}. In our case, the two are in an approximate mutual agreement. 

We have found that the break frequencies in the power spectra, $f_{\rm b}(\nu)$, are compatible with the origin of the emission at $z_\nu$, which are roughly equal to $\beta c/f_{\rm b}(\nu)$ for $\nu<\nu_0$. However, an increasing $f_{\rm b}$ with increasing $\nu$ is also observed for the IR and optical data (see fig.\ 5 of \tet), for which $\nu>\nu_0$. In our jet model, the emission at $\nu>\nu_0$ is the optically-thin synchrotron from the entire part of the jet at $z>z_0$, which implies $z_{\nu>\nu_0}= z_0$. Thus, we expect that the above scaling of $f_{\rm b}\propto z_\nu^{-1}$ no longer holds at $\nu>\nu_0$. Then, the IR/optical variability at high Fourier frequencies may be mostly due to electron energy losses and the re-acceleration (see Section \ref{losses}) rather than due to propagation of some disturbances from $z<z_0$.

As shown in fig.\ 8 of \tet, the optical and IR light curves are tightly correlated, with no measurable lag ($-18^{+30}_{-50}$\,ms), in spite of a relatively large disk contribution in the optical range ($3.7\times 10^5$\,GHz), as shown in Figure \ref{spec}. This shows the the disk contribution is constant on the studied time scales, which is consistent with the rms variability in the optical range reduced with respect to the IR one, see fig.\ 5 in \tet. As shown in \tet, the upper limit on the lag is consistent with the synchrotron energy losses at the magnetic field strength of $\sim 10^4$\,G, which agrees with our determination of $B_0$.

\subsection{Relationship to core shifts}
\label{core_shift}

Time lags are closely related to core shifts, $\Delta\theta$, which are angular displacements of the radio cores, observed at frequencies where the synchrotron emission is partially self-absorbed. They are commonly found in radio-loud active galactic nuclei (e.g., \citealt{Pushkarev12}). The physical cause of the physical displacement along the jet, $z_{\nu_2}-z_{\nu_1}$, is the same for both the core shifts and time lags; only the methods to determine it are different. Using equation (4) in \core and Equation (\ref{Delta_t}), the relationship of $\Delta\theta$ to $\Delta t_{\rm a}$ is
\begin{equation}
\Delta\theta =\frac{\Delta t_{\rm a}\beta c(1+z_{\rm r})\sin i}{D(1-\beta \cos i)}.
\label{core}
\end{equation}
We can then relate $\Delta t_{\rm a}$ to $z_\nu$, $z_0$ and $\nu_0$ using Equations (\ref{Delta_t}--\ref{z0}) and (\ref{z_0}).  

We can estimate $B_0$ using the core-shift method, but only assuming $a=2$, $b=1$, which parameters have been assumed in published core-shift studies, including \core. Using equation (7) of \core and Equation (\ref{core}), we obtain $B_0\approx 1.0\times 10^4$\,G at $p=2$, $\delta=1$ and $\beta_{\rm eq}=1$, in a good agreement with our estimate of Equation (\ref{B0final}). We can also obtain $B_0$ from equation (8) in \core without specifying $\beta_{\rm eq}$. 

\section{Conclusions}
\label{conclusions}

We have based our study on the results of a multiwavelength campaign observing \source when it was close to the peak of its luminous hard spectral state, at $\sim$15\% of the Eddington luminosity. We have used mostly the data published in \tet as well as the IR/optical spectrum from \citet{Rodi21}. Our main conclusions are as follows.

A major model-independent result of our study is the estimate of the distances at which the jet emits below the observed break frequency, based on the time lags between various frequencies. These distances are definitely several orders of magnitude higher than $R_{\rm g}$. By extrapolating the observed approximate correlation of the time lags with the differences between the photon wavelengths, the place where that emission begins can be estimated to be at the distance of several tens of thousands of $R_{\rm g}$ from the BH. This value of that distance agrees with the corresponding finding of \citet{Rodi21}, based on spectral modelling alone.

We then use the classical model of \citet{BK79}, as formulated in detail in later works, to determine the parameters of the jet emitting in the radio-to-optical range. The model assumes the hard inverted spectrum is due to the superposition of locally-emitted spectra that are self-absorbed up to some frequency and then are optically thin. Apart from some details, this is the same model as that used by \tet. The values of the jet parameters obtained by us update those of \tet, which suffered from some errors (see Section \ref{intro} and Appendix \ref{general}). Our analysis is also broader than that of \tet, utilizing constraints from the break frequency and the optically thin part of the spectrum.

By applying the model to the data, we find we cannot uniquely determine the jet bulk Lorentz factor, $\Gamma$, which then needs to be specified as a free parameter. However, it can be constrained from above by the requirement of the jet power being less than the accretion power. It can also be constrained from below by estimating the e$^\pm$ pair production rate in the base of the jet and comparing it to the flux of e$^\pm$ required to account for the observed synchrotron emission. We then use a Bayesian MCMC method to determine all of the jet parameters, and find the most likely range of $1.7\lesssim \Gamma\lesssim 4.1$. We find the jet half-opening angle, $\Theta$ constrained to $\approx$0.6--$2\degr$. The onset of the emission is at $z_0\approx 3$--$4\times 10^{10}$\,cm, where the magnetic field strength is $B_0\approx 0.8$--$1.4\times 10^4$\,G. The total jet power is between $P_{\rm j}\sim 10^{37}$ and $\sim\! 10^{39}$\,erg\,s$^{-1}$. The jet composition is strongly correlated with $P_{\rm j}$, being mostly pairs at the lower limit and mostly e$^-$ and ions at the upper limit. The optical spectral data imply a rather low value of the maximum electron Lorentz factor, of $\gamma_{\rm min}\approx 110$--150.

In order to explain the possible presence of e$^\pm$ pairs in the jet, we calculate the rate of pair production in the jet base in immediate vicinity of the hot accretion flow. We use the measurement of a power-law spectral component extending at least to 2 MeV in the same state of \source. This rate depends on the geometry, see Figure \ref{jet_pairs}, but we find it to be of the same order as the rate of the electron flow through the synchrotron-emitting part of the jet, both being $\sim 10^{40}$--$10^{41}$\,s$^{-1}$. We find this coincidence to be a strong argument for the presence of pairs in the hard-state jet of \source. We have also estimated the contribution of transient accreting BH binaries to the total rate of positron annihilation in the Galaxy and found it to be very small.

We also consider the possibility of the jet power being limited by the power from the rotation of the BH in the presence of magnetically arrested accretion flow. To test it, we calculate the magnetic flux of the jet in the emitting region and that threading the BH. We find them to be very similar, $\sim\! 10^{21}$\,G\,cm$^2$, which remarkable coincidence argues for that scenario. Then, the jet is initially magnetic, Poynting-flux dominated, slow, and not emitting, and accelerates and dissipates its energy only at large distances, in agreement with our finding of the emission being far from the BH.

We find the synchrotron power to be only a small fraction, $\sim\! 10^{-2}$, of the total jet power. On the other hand, the synchrotron power is very similar to either the electron or magnetic powers at the onset of the dissipation, showing that decline of those powers with the distance necessary to explain the observations can be due to the synchrotron emission.

Finally, we show the correspondence between the methods to determine the jet parameters based on time lags and radio core shifts. We give a formula relating the lags and the angular displacements of the radio cores.

\section*{Acknowledgments}
We thank M. B{\"o}ttcher, B. De Marco, P. Gandhi, N. Kylafis, P.-O.\ Petrucci, Th.\ Siegert and A. Veledina for valuable comments and discussions, A. Tchekhovskoy for permission to use his plot of GRMHD simulations, and the referee for valuable comments. We acknowledge support from the Polish National Science Centre under the grants 2015/18/A/ST9/00746 and 2019/35/B/ST9/03944, and from the International Space Science Institute (Bern). Support for this work was provided by NASA through the NASA Hubble Fellowship grant \#HST-HF2-51494.001 awarded by the Space Telescope Science Institute, which is operated by the Association of Universities for Research in Astronomy, Inc., for NASA, under contract NAS5-26555. This paper makes use of the following ALMA data: ADS/JAO.ALMA\#2017.1.01103.T. ALMA is a partnership of ESO (representing its member states), NSF (USA) and NINS (Japan), together with NRC (Canada), MOST and ASIAA (Taiwan), and KASI (Republic of Korea), in cooperation with the Republic of Chile. The Joint ALMA Observatory is operated by ESO, AUI/NRAO and NAOJ. The National Radio Astronomy Observatory is a facility of the National Science Foundation operated under cooperative agreement by Associated Universities, Inc.

\appendix
\section{The general solution}
\label{general}

We provide here the general solution to the equations providing the jet structure, given in Section \ref{power_law}. We first give the solution parametrized by the equipartition parameter, $\beta_{\rm eq}$, but without utilizing the time-lag constraint. This solution is analogous to that given by equations (28--29) in \citet{ZLS12}, which is valid for $a=2$, $b=1$ only. Here, we assume that equipartition holds along the entire emitting jet, i.e., $a=2b$; otherwise, it would be artificial to impose it only at $z_0$. We note that the solutions below are functions of $\gamma_{\rm max}$ (through $f_E$ and $f_N$), while observationally we determine $\nu_{\rm max}$. The relation between the two involves $B_0$, $\Gamma$ and $i$, see Equations (\ref{definitions}) and (\ref{nu_range}). As a consequence, explicit solutions in terms of $\nu_{\rm max}$ would be rather complicated, and we do not provide them.

We determine $B_0$ by setting $n_0$ from Equation (\ref{tau1}) equal to that following from Equation (\ref{betaeq}), and then use $z_\nu=z_0(\nu_0/\nu)^q$ from Equation (\ref{xi_nu}), finding
\begin{align}
&B(z_\nu)=\frac{m_{\rm e}c}{e}
\left(\frac{\pi}{\delta}\right)^{\frac{2+p}{6+p}} \left[\frac{3 c (1+k_{\rm i})(f_E-f_N) \sin i}{\beta_{\rm eq} C_2(p) z_\nu \tan\Theta}\right]^{\frac{2}{6+p}}\nonumber\\ 
&\times \left[2 \nu (1+z_{\rm r})\right]^{\frac{4+p}{6+p}}
\label{B_0eq}
\end{align}
for $z_\nu\geq z_0$, $\nu\leq \nu_0$. Then, $B_0=B(z_0)$. We then substitute $B(z_\nu)$ of Equation (\ref{B_0eq}) in the formula for $F_{\nu}$ in the optically-thick case, Equation (\ref{solution}), which yields for $z_\nu\geq z_0$
\begin{align}
&z_\nu=\frac{1}{\delta^{\frac{4+p}{13+2 p}} \nu} \left[\frac{c (1+k_{\rm i})(f_E-f_N)}{2\pi \beta_{\rm eq} }\right]^{\frac{1}{13+2 p}} \times \nonumber\\
&\left[\frac{C_2(p)}{\sin i}\right]^{\frac{5+p}{13+2 p}}\!\! 
 \left[\frac{F_{\nu} D^2}{m_{\rm e}C_1(p) g_{bp}}\right]^{\frac{6+p}{13+2 p}}\times \label{z_0}\\
&\left(\frac{3}{\pi\tan\Theta}\right)^{\frac{7+p}{13+2 p}} (1+z_{\rm r})^{-\frac{19+3p}{13+2 p}},\nonumber
\end{align}
where $g_{bp}$ follows from Equation (\ref{solution}) for $a=2b$,
\begin{equation}
g_{bp}\equiv \Gamma_{\rm E}\left[\frac{b(p+5)-6}{b(p+6)-2}\right]/(4+b).
\label{gabp}
\end{equation}
Then, $z_0=z_{\nu_0}$, at which $F_{\nu_0}=2 F_0 g_{bp}$, see Equation (\ref{solution})\footnote{Equation (\ref{z_0}) also provides the correct form of equation (5) in \citet{Heinz06} for $p=2$, $b=1$. His equation should be multiplied by $\delta^{1/2}$ factor, which is due to that factor missing in his  equation (1), which should have accounted for the frame transformation from $\nu'$ to $\nu$. That incorrect model formulation was used in \tet.}.

We can then substitute the above $z_\nu\,(\geq z_0)$ into Equation (\ref{B_0eq}),
\begin{align}
&B(\nu)= \nu \left[\frac{3 C_1(p) g_{bp}(1+k_{\rm i})^2 (f_E-f_N)^2\sin^3 i}{C_2(p)^3\beta_{\rm eq}^2 D^2 F_{\nu}\tan\Theta }\right]^{\frac{2}{13+2 p}} \nonumber\\
&\times \frac{\pi^{\frac{7+2 p}{13+2 p}}  2^{\frac{9+2 p}{13+2 p}} c^{\frac{17+2 p}{13+2 p}} m_{\rm e}^{\frac{15+2 p}{13+2 p}} (1+z_{\rm r})^{\frac{15+2 p}{13+2 p}}}{e \delta^{\frac{3+2 p}{13+2 p}}},
\label{B_0f}
\end{align}
and $B_0=B(\nu_0)$. We see that the above solutions are obtained without specifying either $\nu_0$ or $z_0$. Also, the spatial index $b$ enters only in the factor $g_{bp}$, and does not modify the functional dependencies. Equations (\ref{z_0}) and (\ref{B_0f}) are equivalent to equations (28--29) in \citet{ZLS12}, which are for $a=2$, $b=1$, and differ only in the definition of $\beta_{\rm eq}$ and by factors of the order of unity due to a slightly different way of integrating the emission along the jet. 

Next, we can use the independent determination of $z_\nu$ from the time lags, $\Delta t_{\rm a}$. A single measured lag between the frequencies $\nu_2$ and $\nu_1$ determines, via Equation (\ref{Delta_t}), $z_{\nu_2}-z_{\nu_1}$. This can be compared to the prediction using $z_\nu$ of Equation (\ref{z_0}), which yields a constraint between $\Theta$ and $\Gamma$. However, a single measurement of $\Delta t_{\rm a}$ has typically a large error. We can combine them by fitting the relationship between $\Delta t_{\rm a}$ vs.\ $z_{\nu_2}-z_{\nu_1}$. This can be done even when the break frequency, $\nu_0$, is unknown. However, here it is known, and we find it convenient to define $t_0$ by $\Delta t_{\rm a}=t_0(z_{\nu_2}-z_{\nu_1})/z_0$, fitted to a number of measured lags. This then implies $z_0=c t_0 \beta \Gamma\delta$. We can set it equal to that implied by Equation (\ref{z_0}), and solve for $\tan\Theta$ as a function of $\Gamma$,
\begin{align}
&\tan\Theta= \frac{3 \left(\beta\Gamma\nu_0 t_0\right)^{-\frac{13+2 p}{7+p}}}{\pi^{\frac{8+p}{7+ p}} \delta^{\frac{17+3 p}{7+ p}}} \! \left[\frac{(1+k_{\rm i})(f_E-f_N)}{\beta_{\rm eq}}\right]^{\frac{1}{7+ p}} \times \nonumber\\
&\left[\frac{2 C_2(p)}{\sin i }\right]^{\frac{5+p}{7+ p}} \left[\frac{F_0 D^2}{m_{\rm e}c^2 C_1(p) }\right]^{\frac{6+p}{7+ p}}  (1+z_{\rm r})^{-\frac{19+3 p}{7+ p}}\!. 
\label{Theta_f}
\end{align}
Note a relatively strong dependence of $\Theta$ on $t_0$, $\Theta\appropto t_0^{-2}$. We can then insert this $\tan\Theta$ into Equation (\ref{B_0f}) to obtain
\begin{align}
&B_0= \frac{2^{\frac{3+p}{7+ p}} \pi^{\frac{5+p}{7+ p}} (m_{\rm e}\nu_0)^{\frac{9+p}{7+ p}} c^{\frac{11+p}{7+ p}} }{e \delta^{\frac{p-1}{7+ p}}}\times \label{B_0ft}\\
&\left[\frac{\beta\Gamma t_0 C_1(p) (1+k_{\rm i})(f_E-f_N) \sin^2 i}{F_0 \beta_{\rm eq} C_2(p)^2 D^2}\right]^{\frac{2}{7+p}}\!\!(1+z_{\rm r})^{\frac{11+ p}{7+ p}}\!. \nonumber
\end{align}
We determine $n_0$ using the above $B_0$ in Equation (\ref{betaeq}),
\begin{align}
&n_0= \frac{\nu_0^{\frac{18+2p}{7+p}} m_{\rm e}^{\frac{11+p}{7+p}} }{2^{\frac{15+p}{7+p}} \delta^{\frac{2p-2}{7+ p}}e^2}
\left[\frac{c^2 \beta\Gamma t_0 C_1(p)\sin^2 i}{F_0 C_2(p)^2 D^2}\right]^{\frac{4}{7+p}}
\nonumber\\
&\times \left[\frac{\pi \beta_{\rm eq}}{(1+k_{\rm i})(f_E-f_N)}\right]^{\frac{3+ p}{7+ p}} (1+z_{\rm r})^{\frac{22+ 2p}{7+ p}}\!. 
\label{n_0ft}
\end{align}

\bibliography{allbib}{}
\bibliographystyle{aasjournal}

\label{lastpage}
\end{document}